\documentclass{article} \usepackage{epsf} \usepackage{amsmath}
\usepackage{amssymb}
 
\newcommand{\half}{\frac{1}{2}}

\newcommand{\dg}{\delta_\mathrm{g}}
 \newcommand{\dR}{\delta_{R}}
\newcommand{\dL}{\delta_{L}} \newcommand{\dV}{\delta_{V}}
\newcommand{\dA}{\delta_{A}}

\newcommand{\DR}{\frac{i}{\epsilon}\delta_{R}}
\newcommand{\DL}{\frac{i}{\epsilon}\delta_{L}}
\newcommand{\DV}{\frac{i}{\epsilon}\delta_{V}}
\newcommand{\DA}{\frac{i}{\epsilon}\delta_{A}}

\newcommand{\hPR}{\hat{P}_{R}} \newcommand{\hcPR}{\hat{{\cal P}}_{R}}
\newcommand{\hPL}{\hat{P}_{L}} \newcommand{\hcPL}{\hat{{\cal P}}_{L}}
\newcommand{\PR}{{P}_{R}} \newcommand{\PL}{{P}_{L}}
\newcommand{\cO}{{\cal O}} \newcommand{\g}{{\gamma_5}}
\newcommand{\gh}{{\hat{\gamma}_5}} \newcommand{\Gh}{{\hat{\Gamma}_5}}
\newcommand{\RRi}{\frac{1}{2R}}

\newcommand{\fm}{\,{\rm fm}}
\newcommand{\MeV}{\,{\rm MeV}}
\newcommand{\GeV}{\,{\rm GeV}}
\newcommand{\psibar}{\overline{\psi}}
\newcommand{\Psibar}{\overline{\Psi}}
\newcommand{\bsplit}{\begin{equation} \begin{split}}
    \newcommand{\esplit}{\end{split} \end{equation}}

\newcommand{\newappendix}[1]{ \vspace{15mm} \pagebreak[3]
  \addtocounter{section}{1} \setcounter{equation}{0}
  \setcounter{subsection}{0} \setcounter{footnote}{0}
  \renewcommand{\theequation}{\Alph{section}.\arabic{equation}}
\begin{flushleft}
  {\Large\bf Appendix \Alph{section}. #1}
\end{flushleft}
\nopagebreak \medskip \nopagebreak}

\begin{document}
\thispagestyle{empty} \parskip=12pt \raggedbottom
 
\def\mytoday#1{{ } \ifcase\month \or January\or February\or March\or
  April\or May\or June\or July\or August\or September\or October\or
  November\or December\fi
  \space \number\year}
\noindent
\hspace*{6cm} BUTP/02/5, UCSD/PTH 02-04\\
\vspace*{1cm}
\begin{center}
  {\LARGE Testing the fixed-point QCD action and the construction of
    chiral currents }
 
  \vspace{1cm} P. Hasenfratz, S. Hauswirth, T. J\"org and F.
  Niedermayer\footnote{On leave from the Institute of Theoretical
    Physics, E\"otv\"os University, Budapest}
  \\
  Institute for Theoretical Physics \\
  University of Bern \\
  Sidlerstrasse 5, CH-3012 Bern, Switzerland
  
  \vspace{0.5cm}
  K. Holland\\
  Department of Physics \\
  University of California at San Diego\\
  9500 Gilman Drive, La Jolla CA 92093, USA
  
  \vspace{0.5cm}
  
  \nopagebreak[4]
 
\begin{abstract}
  We present the first set of quenched QCD measurements using the
  recently parametrized fixed-point Dirac operator $D^{\rm FP}$. We
  also give a general and practical construction of covariant
  densities and conserved currents for chiral lattice actions. The
  measurements include (a) hadron spectroscopy, (b) corrections of
  small chiral deviations, (c) the renormalized quark condensate from
  finite-size scaling and, independently, spectroscopy, (d) the
  topological susceptibility, (e) small eigenvalue distributions and
  random matrix theory, and (f) local chirality of near-zero modes and
  instanton-dominance.
\end{abstract}
 
\end{center}
\eject

\section{Introduction}
The only way to study a quantum field theory non-perturbatively from
first principles is via lattice regularization. For a
strongly-interacting theory like QCD, many essential features are
non-perturbative. Over the course of more than two decades, lattice
QCD has determined, with varying degrees of accuracy, the hadron mass
spectra, quark masses, the strong interaction coupling constant,
low-energy constants, inter-particle potentials, the phase structure
of the theory, decay constants, matrix elements and many other
quantities \cite{Kaneko:2001ux}.  To make lattice QCD into a precise
science requires detailed study of the systematic error introduced by
the lattice discretization. As observed recently in a study of the
$2d~O(3)$ non-linear sigma model \cite{Hasenfratz:2000sa}, assumptions
about how to extrapolate lattice results to the continuum may be
wrong.  Even for very accurate lattice measurements, large deviations
as the lattice spacing $a$ is varied could mean large uncertainty in
continuum quantities, due to the extrapolation ansatz.  To reduce such
sensitivity, many groups now use e.g.~${\cal O}(a)$-improved lattice
actions to remove the leading order lattice artifacts. However, it may
still be necessary to go to very fine resolution before the continuum
extrapolation can be done confidently.

A separate issue is that, for a long time, it was thought impossible
to have chiral lattice fermions \cite{Nielsen:1980rz}.  This prevented
a clean study of the chiral aspects of QCD, such as spontaneous
symmetry breaking, and caused several technical headaches, such as
fine-tuning the bare quark mass to make the pions light, exceptional
configurations, mixing of operators with different chiral
representations and additional renormalization factors. The question
was also raised whether a chiral theory, such as the Standard Model,
could in principle be defined non-perturbatively.  The domain wall
fermions \cite{Kaplan:1992bt}, the related overlap construction
\cite{Narayanan:ss} and the fixed-point (FP) action
\cite{Hasenfratz:1993sp} are the known examples for lattice
regularized fermions with chiral symmetry. The common feature behind
these rather different constructions is the Ginsparg-Wilson (GW)
relation \cite{Ginsparg:1981bj} which is satisfied by all these
constructions \cite{Hasenfratz:1997aa}.  Actually, one might construct
a Dirac operator using directly the GW relation and the requirement of
locality \cite{Gattringer:2000ja}.  The GW relation not only implies
all the physical consequences of chiral symmetry
\cite{Hasenfratz:1998ri,Hasenfratz:1998bb}, but also the existence of
an exact symmetry transformation \cite{Luscher:1998pq}. This has
caused an explosion of interest in the last few years in theory
\cite{Niedermayer:1998bi} and in using different methods to explore
the chiral behavior of quenched QCD (no sea quarks), which is now
quickly maturing. With the standardly used algorithms, it is
impossible to use chiral lattice actions for simulations of full QCD
(with sea quarks), as these actions have a complicated structure
\cite{Jansen:2001fn}. This is the major obstacle for the future use of
chiral actions. For a recent promising development see, however,
ref.~\cite{Knechtli:2000ku}.

The fixed-point method is motivated by renormalization group
properties of lattice field theories \cite{Hasenfratz:gu}.  Instead of
trying to remove lattice artifacts order by order in the lattice
spacing $a$, a fixed-point (FP) action is designed to be less
sensitive to the discretization even at large lattice spacings.  In
fact, for some quantities, a FP action is entirely blind to the
discretization. An exact FP QCD action has exact chiral symmetry and
offers a completely new approach to examine the chiral properties of
the theory. Exact and approximate fixed-point actions have been
constructed and tested for a number of models, including pure
Yang-Mills theory, and in general have shown very good scaling
behavior \cite{Rufenacht:2001pi,Blatter:1994hy}.  
The construction and initial tests of an approximate fixed-point 
Dirac operator $D^{\rm FP}$ are described in
\cite{Hasenfratz:2000xz}. 
Like other chiral actions, fixed-point actions are more costly 
to use in simulations than standard actions.  However, FP actions 
are designed to have much reduced lattice artifacts and, if chiral 
actions are the correct long-term approach for lattice QCD, 
FP actions might be the optimal choice.

In this paper, we present the results of a preliminary study of an
approximate FP QCD action, to test if it is feasible to use such an
action in simulations. A larger systematic study, which is running
already as part of the program of the Bern-Graz-Regensburg
Collaboration, is required to examine the lattice spacing dependence
of measurements using this action (see ref.~\cite{Hauswirth2002}).
As well as these initial
measurements, we also give a practical method to construct and use
conserved currents in simulations with chiral lattice actions.  Many
groups now use chiral actions, but non-conserved currents and/or
non-covariant densities and so lose part of the advantages offered by
a chiral formulation.

This paper is a detailed description of the work summarized in
\cite{Hasenfratz:2001hr,Hasenfratz:2001qp}.  The paper is organised as
follows. In Section 2, we examine how small chiral deviations of the
action can be corrected. Hadron spectroscopy measurements with the
parametrized FP Dirac operator are described in Section 3, giving the
pion, rho and nucleon masses, the speed of light, the remnant additive
quark mass renormalization and an indication of the presence of
quenched chiral logarithms.  In Section 4, we use the overlap-improved
FP Dirac operator to measure the meson spectrum, the renormalized
quark condensate via finite-size scaling, the quenched topological
susceptibility, investigate random matrix theory and small eigenvalue
distributions on small lattice volumes, and examine
instanton-dominance of near-zero modes of the Dirac operator.  The
construction of covariant currents and densities and a discussion of
Ward identities for chiral lattice actions are given in Section 5,
followed by our conclusions. In the Appendix we collect some useful
identities which are implied by the Ginsparg-Wilson relation.

\section{{}Chiral behavior of ${\mathbf D^{\rm FP}}$}
A Dirac operator $D$ which satisfies exactly the Ginsparg-Wilson
relation
\begin{equation}
\label{gw}
\g D + D \g = D \g 2R D \,,
\end{equation}
where $R$ is any local function and trivial in Dirac space, has exact
chiral symmetry even at non-zero lattice spacing $a$. Such a Dirac
operator has infinitely many couplings \cite{Horvath:1998dd}.
Simulations are only feasible with a finite number of couplings, hence
the chiral symmetry is only approximate.  What needs to be examined is
how good the approximation is.  The exact fixed-point Dirac operator
satisfies exactly the Ginsparg-Wilson relation. The parametrized
fixed-point Dirac operator $D^{\rm FP}$, which we use in simulations,
is an approximate solution to the QCD fixed-point equations. The
construction and initial tests of $D^{\rm FP}$ are described in
\cite{Hasenfratz:2001hr,Hasenfratz:2001qp}.  Here, we test the quality
of the chiral symmetry of this operator.

\begin{figure}[htb]
\begin{center}
  \vskip -3mm \leavevmode \epsfxsize=100mm \epsfysize=100mm
  \epsfbox{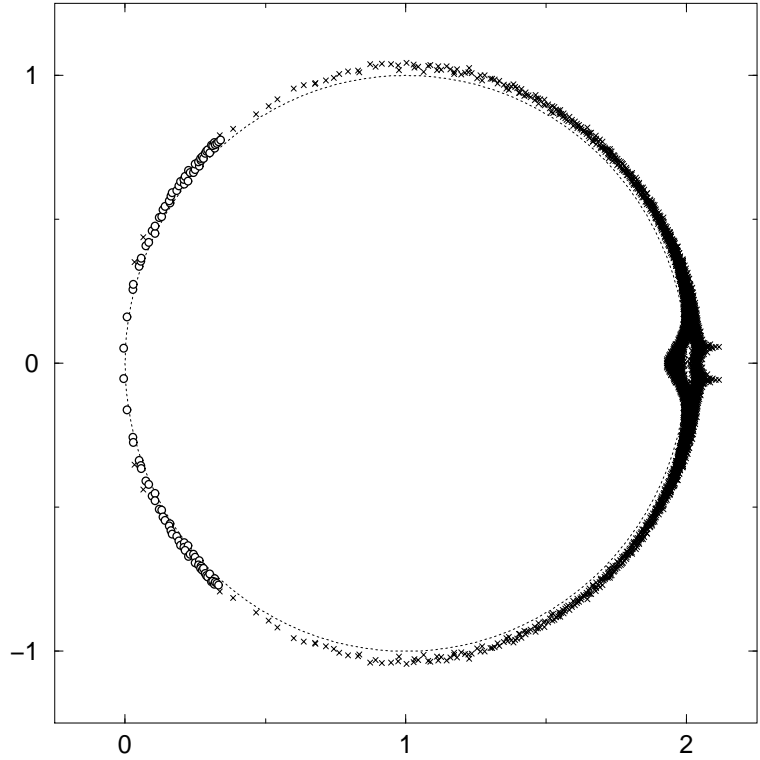} \vskip -6mm
\end{center}
\caption{{}\label{fig:GWcircle} The full eigenvalue spectrum of 
  $D^{\rm FP}$ for a $S^{\rm FP}_g$ gauge configuration of 
  volume $4^4$ (crosses) and the smallest 100 eigenvalues on 
  a $8^4$ volume (circles) at a lattice spacing $0.16\fm$. }
\end{figure}

As will be discussed in Sec.~5, the function $2R$ can be absorbed into
the definition of $D$, so for simplicity, we set $2R=1$ in most of the
following equations. The Ginsparg-Wilson relation can be re-written as
\begin{equation}
D + D^{\dagger} = D^{\dagger} D,
\end{equation}
using the property $\g D \g = D^{\dagger}$. If $D$
satisfies this relation, its complex eigenvalues lie on the circle of
radius 1 and centre at (1,0) in the complex plane. In
Fig.~\ref{fig:GWcircle}, we plot the complex eigenvalues of 
$D^{\rm FP}$ for fixed-point gauge action $S^{\rm FP}_g$
configurations. 
The lattice spacing can be determined via the Sommer parameter $r_0$ of
the static quark anti-quark potential. To orient the reader, the bare
coupling of the gauge action $\beta = 6/g^2 = 3.0$ and $3.2$
correspond to lattice spacings 
$a \simeq 0.16\fm$ and $a \simeq 0.13\fm$, respectively.  The figure
contains all the eigenvalues for $4^4$ gauge configurations and those
closest to the origin for $8^4$ configurations.  It's clear that the
breaking of the Ginsparg-Wilson relation is small over the entire
eigenvalue spectrum even though the lattice spacing is quite coarse,
and that the chiral symmetry remains very good even for larger
volumes.  For comparison, the eigenvalues of the Wilson Dirac operator
$D^{\rm W}$ at a similar lattice spacing are very far removed from the
Ginsparg-Wilson circle \cite{Bietenholz:2000iy}.

\begin{figure}[htb]
\begin{center}
  \vskip -3mm \leavevmode \epsfxsize=85mm \epsfbox{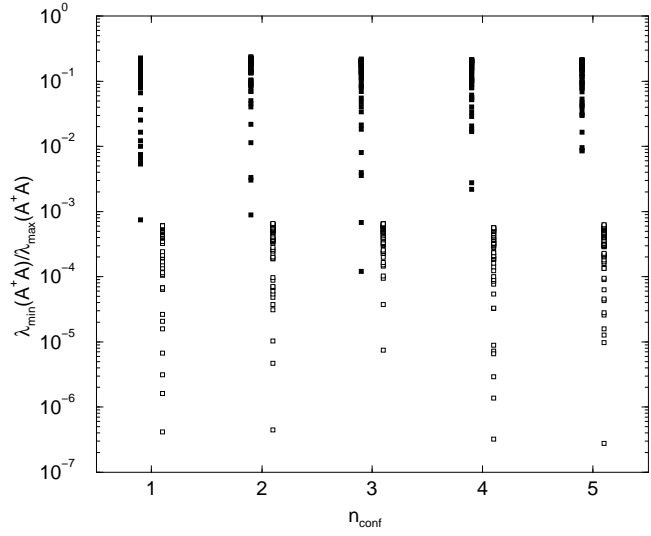}
  \vskip -8mm
\end{center}
\caption{{}\label{fig:AdaggerA} Ratio of the 50 smallest 
  $A^{\dagger} A$ eigenvalues to the largest eigenvalue using 
  $D^{\rm FP}$ (filled squares) and $D^{\rm Wilson}$ (open squares) 
  for 5 different $S^{\rm FP}_g$ gauge configurations of volume 
  $12^4$ at a lattice spacing $0.16\fm$.}

\end{figure}

Defining the operator $A = 1 - D$, the Ginsparg-Wilson relation is
equivalent to the requirement that $A$ be unitary i.e.~$A^{\dagger} A
= 1$. We can measure how much this requirement is broken by finding
the smallest and largest eigenvalues of $A^{\dagger} A$. In
Fig.~\ref{fig:AdaggerA}, we plot the ratio of the 50 smallest
eigenvalues of $A^{\dagger} A$ to the largest eigenvalue for a number
of $S^{\rm FP}_g$ gauge configurations with a lattice spacing 
$0.16 \fm$. 
The largest eigenvalue is approximately 1.5 and 38 for $D^{\rm FP}$ 
and $D^{\rm W}$ respectively. We see that the
fixed-point operator is much closer than the Wilson operator to
satisfying the unitarity condition.  Also, as the volume increases,
it's more likely that using $D^{\rm FP}$ produces a few very small
eigenvalues, but the vast bulk of the $A^{\dagger} A$ spectrum is
close to 1.

For some measurements, it's necessary that the remnant chiral symmetry
breaking be very small. Given any suitable Dirac operator $D_0$ as
input, the overlap Dirac operator \cite{Narayanan:ss}
\begin{equation}
  \label{eq:overlap}
  D_{\rm ov} = 1 - A/\sqrt{A^{\dagger} A}, \hspace{1cm} A = 1 + s - D_0
\end{equation}
is an exact solution of the Ginsparg-Wilson relation. The real
parameter $s$ can be used to optimize the convergence of
approximations or the locality of the resulting overlap Dirac
operator.  The nice feature of this operator is that it's an explicit
construction, unlike the fixed-point operator which is a solution to a
set of equations. If the input operator $D_0$ is already chiral, then
$A^{\dagger} A = 1$ and hence $D_{\rm ov}=D_0$. Most simulations using
$D_{\rm ov}$ have taken $D^{\rm W}$ as the input operator, which has
severely broken chiral symmetry, as $A^{\dagger} A$ is far from 1 and
hence computing $1/\sqrt{A^{\dagger} A}$ is a very expensive numerical
problem. A measure of this difficulty is the condition number of
$A^{\dagger} A$, the ratio of its smallest to largest eigenvalue.
Also, the overlap operator may inherit the large lattice artifacts of
$D^{\rm W}$. Alternatively, taking $D^{\rm FP}$ as the input operator,
$A^{\dagger} A$ is close to 1 and $1/\sqrt{A^{\dagger} A}$ is easier
to evaluate --- the improvement in the condition number is clear in
Fig.~\ref{fig:AdaggerA}.  The overlap construction should only bring
small corrections to $D^{\rm FP}$. This way, the residual chiral
symmetry breaking can be removed without destroying the very important
fixed-point properties.

\begin{figure}[htb]
\begin{center}
  \vskip -1mm \leavevmode \epsfxsize=80mm \epsfbox{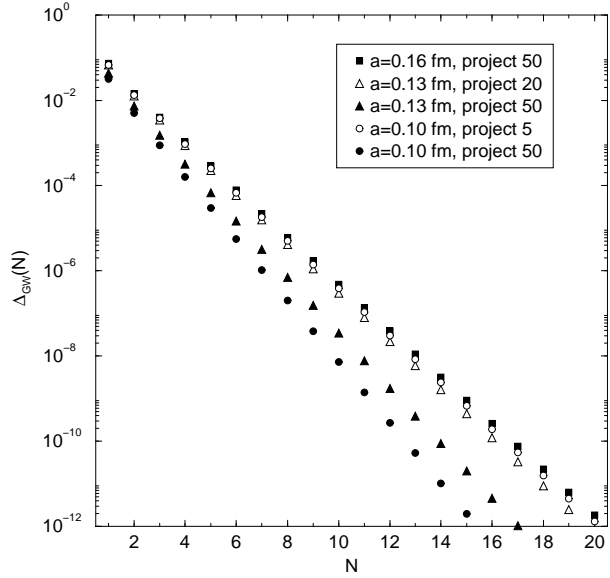}
  \vskip -8mm
\end{center}
\caption{{}\label{fig:GWbreak_vec} Breaking of the Ginsparg-Wilson 
  relation as measured by $\Delta_{\rm GW}(N)$ on $10^4$ lattices 
  at different lattice spacings and different number of exactly 
  projected $A^{\dagger} A$ eigenvalues. 
  Keeping the number of exactly projected $A^{\dagger} A$
  eigenvalues constant, the exponential fall-off is steeper for
  smaller lattice spacings. If the number of exactly projected
  $A^{\dagger} A$ eigenvalues is adjusted such that the range of the
  not exactly treated $A^{\dagger} A$ eigenvalues is approximately the
  same for the different lattice spacings, the fall-off is roughly
  equal in all cases, which illustrates that the approximation is
  governed by the ratio 
  $\lambda_{\rm min}(A^{\dagger} A)/\lambda_{\rm max}(A^{\dagger} A)$.}
\end{figure}

As the chiral symmetry is already well approximated by $D^{\rm FP}$,
we use a Legendre polynomial approximation of $1/\sqrt{A^{\dagger} A}$
to construct the overlap-improved operator $D_{\rm ov}^{\rm FP}$.  As
$A^{\dagger} A$ may occasionally have a few very small eigenvalues, we
always project out the smallest 10 to 100 eigenvalues, which are
treated exactly. One measure of the remnant chiral symmetry breaking
is the vector norm 
$\Delta_{\rm GW}=||(D+D^{\dagger} - D^{\dagger}D) v||$, 
where $v$ is a unit vector of random entries. 
If $D$ solves the Ginsparg-Wilson relation, $\Delta_{\rm GW}$ vanishes.
In Fig.~\ref{fig:GWbreak_vec}, we plot $\Delta_{\rm GW}(N)$ versus the
order $N$ of the polynomial used to approximate 
$1/\sqrt{A^{\dagger} A}$ for gauge configurations of volume 
$10^4$ at different lattice spacings ($0.16\fm$, $0.13\fm$, $0.10\fm$).
The chiral symmetry breaking falls off exponentially as 
the polynomial order increases.  
The rate of the fall off varies depending on the range of
the $A^{\dagger} A$ eigenvalues which are not projected out.
Projecting out the same number of $A^{\dagger}A$ eigenvalues for the
different lattice spacings, the rate of the fall off is the largest
for the smallest lattice spacing. However, choosing the number of
projected $A^{\dagger} A$ eigenvalues such that ratio $\lambda_{\rm
  min}(A^{\dagger} A)/\lambda_{\rm max}(A^{\dagger} A)$ for the
remaining $A^{\dagger} A$ eigenvalues is approximately the same for
the three different lattice spacings, the rate of the fall off is
roughly equal for all the lattice spacings, which illustrates that the
convergence rate is indeed governed by the ratio $\lambda_{\rm
  min}(A^{\dagger} A)/\lambda_{\rm max}(A^{\dagger} A)$.

\begin{figure}[htb]
\begin{center}
  \vskip -3mm \leavevmode \epsfxsize=100mm
  \epsfbox{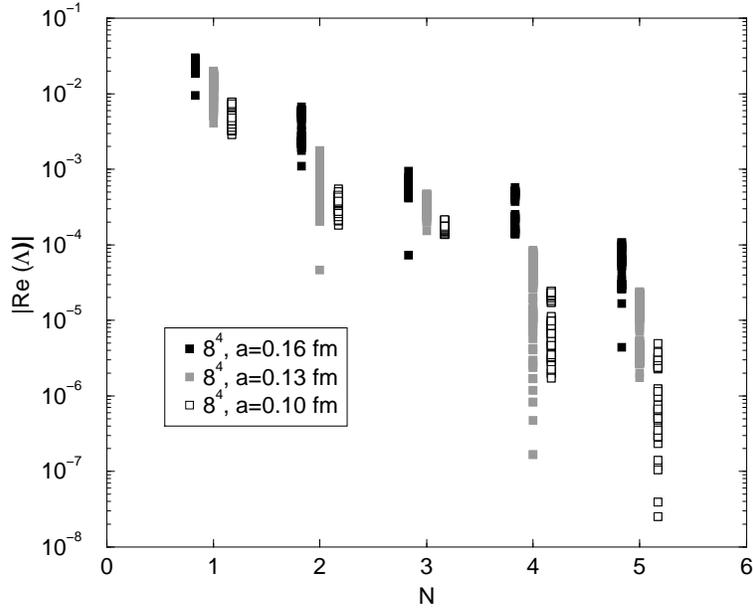} \vskip -6mm
\end{center}
\caption{{}\label{fig:GWbreak_eig} The dependence of the deviation of the 
  $D_{\rm ov}^{\rm FP}$ eigenvalues from the Ginsparg-Wilson circle 
  on the order of the polynomial approximation,
  as measured by $|{\rm Re}(\Lambda)|$.}
\end{figure}

If the eigenvalues $\lambda_n$ of $D$ lie exactly on the
Ginsparg-Wilson circle, then $\Lambda_n = \lambda_n/(1-\lambda_n/2)$
are purely imaginary i.e.~the eigenvalues are projected from the
circle onto the imaginary axis. In Fig.~\ref{fig:GWbreak_eig}, we plot
$|{\rm Re}(\Lambda_n)|$ versus the order of polynomial used to
approximate $1/\sqrt{A^{\dagger} A}$, for gauge configurations with
lattice spacings $0.16, 0.13$ and $0.10\fm$. As the order
increases, $|{\rm Re}(\Lambda_n)|$ becomes exponentially small as the
eigenvalues $\lambda_n$ lie closer and closer to the Ginsparg-Wilson
circle, with the most rapid decrease at the finest lattice spacing.

\begin{figure}[htb]
\begin{center}
  \vskip 0mm \leavevmode \epsfxsize=95mm \epsfbox{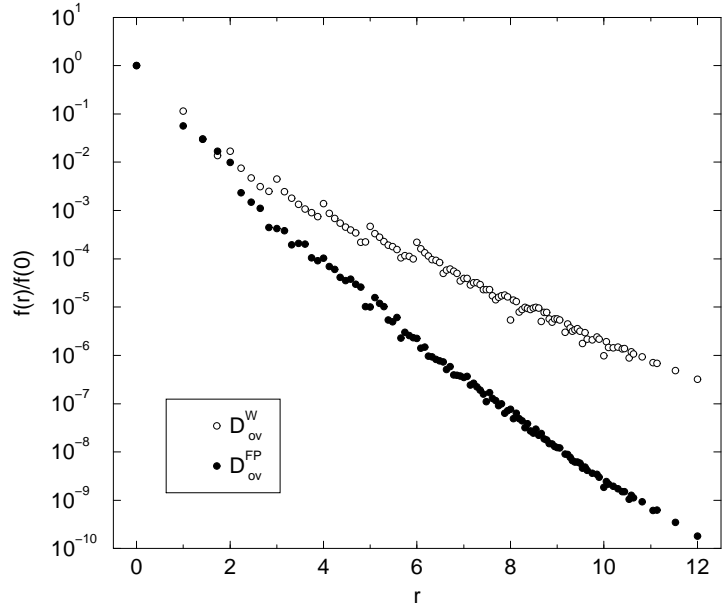} \vskip
  -6mm
\end{center}
\caption{{}\label{fig:locality} The locality of $D_{\rm ov}$, 
  as measured by the expectation value $f(r)/f(0)$ defined in 
  Eq.~\eqref{eq:locality}, using $D^{\rm FP}$ and $D^{\rm Wilson}$ 
  as input on $12^4$ lattices at $a=0.16\fm$.}
\end{figure}

A lattice action must be local for its continuum predictions to be
universal i.e.~independent of the choice of lattice discretization.
Locality on the lattice means that fields at large separation have an
exponentially small coupling. The exponential decay of the action
should be faster than that of the correlation functions. This issue
was first addressed for the overlap operator $D_{\rm ov}$ in
\cite{Hernandez:1998et}, where a free parameter was varied to maximize
the exponential fall-off.  One measure of locality is
\begin{equation}
  \label{eq:locality}
  f(r) = \max_y \{ ||D_{\rm ov} v||, ||y-x||=r \},
\end{equation}
where $v$ is vector with a point source at $x$ and $r$ is the square
norm. In Fig.~\ref{fig:locality}, we plot the expectation value of
$f(r)/f(0)$ versus $r$ for different overlap operators on $12^4$
volumes at a lattice spacing of $0.16\fm$.  Compared with other
tests of locality of $D_{\rm ov}$ using $D^{\rm W}$ as input
\cite{Hernandez:1998et,Bietenholz:2001nu}, the locality is
significantly improved if the overlap operator is constructed using
$D^{\rm FP}$, with a faster exponential decay. The exponent $\nu$
describing the decay at large separation $r$ is $\nu = 0.94$ for
$D^{\rm W}_{\rm ov}$, whereas it is $\nu = 1.60$ for $D^{\rm FP}_{\rm
  ov}$. In order to optimize the locality of $D^{\rm W}_{\rm ov}$ the
parameter $s$ in Eq.~\eqref{eq:overlap} has to be tuned and we find
$s=0.6$ as optimal value at this lattice spacing. In contrast to
$D^{\rm W}_{\rm ov}$ there is no tuning of the parameter $s$ needed
for $D^{\rm FP}_{\rm ov}$ to get the (approximately) optimal locality,
i.e.we can use~$s=0.0$ for $D^{\rm FP}_{\rm ov}$.

From these tests, we see that the chiral symmetry is well approximated
by $D^{\rm FP}$ and that the residual breaking can be removed using
overlap-improvement. Only a relatively low order polynomial
approximation is required to construct $D_{\rm ov}^{\rm FP}$, which is
essential as $D^{\rm FP}$ is more expensive to use than $D^{\rm W}$.
The parametrized FP operator has 9 times more offsets and the
computational cost per offset is also higher since $D^{\rm FP}$
includes all elements of the Clifford algebra.  If, in a matrix-vector
multiplication, for $D^{\rm W}$ the time needed for Dirac matrix
multiplications is neglected, then $D^{\rm FP}$ is a factor of
$\approx 4$/offset more expensive leading to an estimated relative
overhead of $\approx 36$. Note, however, that this number may vary
quite a bit depending on the underlying computer architecture, as one
of the main issues in present lattice simulations is rather fast
memory access and fast communication than fast floating point units
\cite{Gottlieb:2001sy}.

\section{Spectroscopy}
Measuring the mass spectrum is a standard benchmark for using a
particular lattice action. Simulations with improved Wilson and
staggered fermions in full QCD are now very advanced in reaching the
physical theory.  Mass spectroscopy in quenched QCD with chiral
fermions is quickly maturing and is a very good test of the chiral
behavior of the theory. In this preliminary study, we test the
behavior of $D^{\rm FP}$, for example, how intact are the fixed-point
properties, how large is the additive quark mass renormalization, how
small an $m_{\pi}/m_{\rho}$ ratio can be attained and do we observe
quenched chiral logarithms.

\subsection{Simulation parameters}
The FP gauge action we use has previously been studied in detail
\cite{Rufenacht:2001pi}.  To remind the reader, the bare coupling of
the gauge action $\beta = 6/g^2 = 3.0$ and $3.2$ correspond to lattice
spacings $a \simeq 0.16\fm$ and $a \simeq 0.13\fm$ respectively, 
as determined from the Sommer parameter $r_0$ of the static 
quark anti-quark potential \cite{Sommer:1993ce}. 
Lattice volumes of $6^3 \times 16$ and
$9^3\times 24$ are compared to investigate the volume dependence of
zero mode effects. We use gaussian smeared sources and point sinks.
Configurations were fixed to Landau gauge using the Los Alamos
algorithm with stochastic overrelaxation \cite{DeForcrand:1989aa}.
With the parametrized FP Dirac operator $D^{\rm FP}$, we simulate at
input quark masses $ma$ ranging from 0.016 to 0.23.  The smallest
quark mass corresponds to $m_\pi/m_\rho = 0.30(3)$.  It has to be
stressed that we introduce the mass in the most simple manner by
defining\footnote{{}The non-covariant form $D^{\rm FP} + m$ for the
  massive Dirac operator was used for the results on the $6^3\times
  16$ lattice and the finite-momentum calculations on the $9^3\times
  24$ lattice} 
$D(m) = D^{\rm FP} + m$ or with the covariant scalar density 
in Eq.~\eqref{Act_mod} $D(m) = (1-m/2)D^{\rm FP} + m(2R)^{-1}$, 
whereas the fixed-point Dirac operator for non-zero mass
would in fact need a reparametrization by iteratively solving the RG
transformations for every mass value as we did for the zero-mass Dirac
operator $D^{\rm FP}$. Therefore our parametrization deviates more and
more from the classical renormalized trajectory for larger masses. 
A multimass BiCGstab algorithm \cite{Jegerlehner:1997rn} is used to
invert the Dirac operator at all masses simultaneously.  Errors are
estimated with bootstrap resampling of the hadron correlators, which
were symmetrized around $t=T/2$ to increase statistics. A correlated
fit to the measured correlators is performed in the interval
$[t_{\rm min},t_{\rm max}]$, where the minimum time $t_{\rm min}$ 
of the fit range is determined by two criteria: i) the effective 
mass starts to show a reasonable plateau, 
ii) the value of $\chi^2$ per degree of freedom in the fit 
($\chi^2/df$) starts to show a plateau and is of order 1. 
The maximum time $t_{\rm max}$ is generally set to $T/2$ in the mass
measurements, while for the finite momentum measurements it is reduced
according to the length of the plateau.

\begin{table}[htb]
\begin{center}
\begin{tabular}{|c|c|c|c|c|c|} \hline
fermion action & lattice size & $\beta$ & $L_s$ & $N_{\rm conf}$ & $ma$ \\ 
\hline
parametrized FP & $6^3 \times 16$ & 3.0 & 1.0 fm & 96 & $0.02,\dots,0.20$ \\
parametrized FP & $9^3\times 24$ & 3.0 & 1.5 fm & 70 & $0.016,\dots,0.23$ \\
overlap FP & $9^3\times 24$ & 3.0 & 1.5 fm & 28 & $0.012,\dots,0.23$ \\ 
overlap FP & $9^3\times 24$ & 3.2 & 1.2 fm & 32 & $0.012,\dots,0.23$ \\ 
\hline
\end{tabular}
\end{center}
\caption{{}Simulation parameters for hadron spectroscopy.}
\end{table}

\begin{figure}[htb]
\begin{center}
  \vskip -3mm \leavevmode \epsfxsize=90mm
  \epsfbox{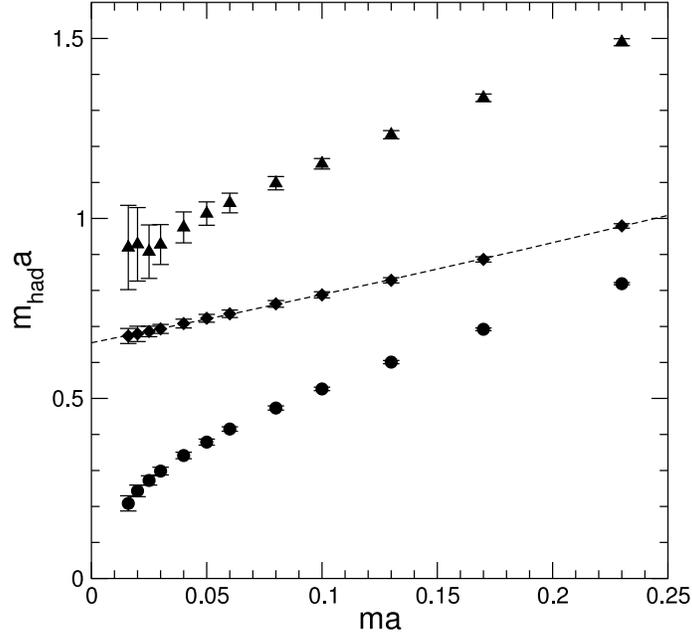}
  \vskip -6mm
\end{center}
\caption{{}\label{fig:fp_spectrum} $\pi$, $\rho$ and N masses 
  for the parametrized FP Dirac operator $D^{\rm FP}$ at $\beta=3.0$ 
  from 70 gauge configurations with lattice size $9^3\times 24$. 
  The dashed line shows a quadratic fit to the rho meson, 
  leading to a lattice spacing of $a\simeq 0.17\fm$.}

\end{figure}

\begin{figure}[htb]
\begin{center}
  \leavevmode \epsfxsize=85mm \epsfbox{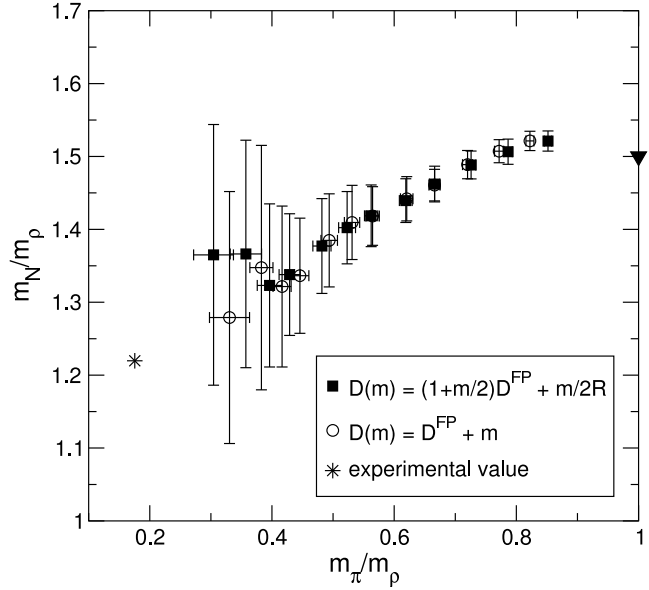}
  \vskip -6mm
\end{center}
\caption{{}\label{fig:fp_edinburgh} Edinburgh plot for 
  the parametrized FP Dirac operator on $9^3\times 24$ at $\beta=3.0$ 
  for the naive and covariant mass definition.}
\end{figure}

Fig.~\ref{fig:fp_spectrum} shows the masses of the pseudoscalar and
vector meson and the nucleon measured with the parametrized FP Dirac
operator on the $9^3\times 24$ lattice. The inversion of the Dirac
operator converged for all configurations within reasonable time,
although for one configuration the number of inversions was about
three times larger than the typical value of $\approx 250$. This
indicates that the residual chiral symmetry breaking leads to
fluctuations on the order of our smallest quark mass for the real
eigenvalues of the Dirac operator, and therefore it would not be wise
to go to smaller masses than $ma=0.016$ with the parametrized FP
operator at this lattice spacing and volume. An extrapolation of the
rho mass to the physical value gives a lattice spacing of $a\simeq
0.17\fm$, which is slightly higher than that obtained from the
Sommer parameter. The sign and size of this deviation is consistent
with earlier findings \cite{AliKhan:2001tx}.
Fig.~\ref{fig:fp_edinburgh} is an Edinburgh plot from this data.

\subsection{The pion mass in the chiral limit}
A lattice calculation of the pion mass in the limit of small quark
mass with the FP action is interesting for several reasons: first, it
serves as a quantitative check of how chiral our action actually is.
For an exactly chiral fermion action like the overlap or the exact
fixed-point action, there is no additive renormalization of the quark
mass. But does the approximation we make by parametrizing the FP
action introduce a residual mass, and can we measure it?  Second, at
the lattice volumes we are working with, one expects to see
topological finite-volume effects proportional to $1/(m_q^2\sqrt{V})$
and $1/(m_q\sqrt{V})$ \cite{Blum:2000kn} which complicate a reliable
pion mass measurement at small quark masses. It is therefore crucial
to check whether these effects are under control. Third, having an
action with good chiral properties at hand, one can check for the
appearance of logarithmic terms in the chiral extrapolation of the
squared pion mass.

\begin{table}[htb]
\begin{center}
\begin{tabular}{|l|lc|lc|lc|} \hline

$ma$ & $m_\pi$(P) & $\chi^2/df$ & $m_\pi$(A) & $\chi^2/df$ &
  $m_\pi$(P-S) & $\chi^2/df$ \\ \hline
0.02 & 0.394(41) & 0.77        & 0.345(62) & 1.55    & 0.273(33) & 1.19 \\
0.03 & 0.398(25) & 1.01        & 0.381(30) & 0.62    & 0.341(32) & 0.72 \\
0.04 & 0.416(19) & 0.92        & 0.403(21) & 0.64    & 0.381(28) & 0.47 \\
0.05 & 0.439(16) & 0.87        & 0.428(19) & 0.80    & 0.418(22) & 0.33 \\
0.06 & 0.463(14) & 0.90        & 0.451(17) & 0.98    & 0.451(21) & 0.25 \\
0.08 & 0.502(15) & 0.91        & 0.498(14) & 1.14    & 0.512(18) & 0.19 \\
0.10 & 0.546(13) & 0.74        & 0.541(12) & 1.10    & 0.566(16) & 0.22 \\
0.13 & 0.611(12) & 0.53        & 0.603(10) & 0.93    & 0.638(16) & 0.37 \\
0.16 & 0.672(11) & 0.38        & 0.663(10) & 0.78    & 0.703(15) & 0.64 \\
0.20 & 0.748(10) & 0.26        & 0.737(08) & 0.64    & 0.783(13) & 1.11
\\ \hline 

\end{tabular}
\end{center}
 \caption{{}\label{tab:m_pi_6x16} Pion masses for parametrized FP
  fermions on $6^3\times 16$ lattice from three different pion
  operators. The quark mass is introduced by
   $D^{\rm FP}(m) = D^{\rm FP}+m$.}
\end{table}

We use three different operators to extract the pion mass: the
pseudoscalar $P = \Psibar\g\Psi$, the fourth component of the
axial vector $A_4=\Psibar\gamma_4\g\Psi$ and the difference of
pseudoscalar and scalar $S=\Psibar\Psi$.  In the quenched theory, the
pseudoscalar correlator $\langle P(x)P(0)\rangle$ is contaminated by
the topological finite size effects from configurations with
non-trivial topology, as mentioned above. The same special finite size
contributions appear in the scalar correlator, so by building the
difference of the pseudoscalar and scalar correlator $\langle P(x)P(0)
\rangle - \langle S(x)S(0) \rangle $, this artificial effect should
cancel, although if the action is not exactly chiral, the cancellation
is also not exact. For larger quark masses, the P-S correlator is
contaminated by effects from the scalar part and is expected to
deviate from the pion. For the axial vector correlator $\langle
A_4(x)A_4(0)\rangle$, the divergent contribution is expected to be
partially suppressed \cite{Blum:2000kn,DeGrand:2000gq}.

\begin{table}[htb]
\begin{center}
\begin{tabular}{|l|lc|lc|lc|} \hline
$ma$ & $m_\pi$(P) & $\chi^2/df$ & $m_\pi$(A) & $\chi^2/df$ &
  $m_\pi$(P-S) & $\chi^2/df$ \\ \hline
0.016 & 0.267(20) & 1.25 & 0.223(15) & 0.53      & 0.209(21) & 0.64 \\
0.02 & 0.274(15) & 1.18  & 0.256(14) & 0.72      & 0.243(16) & 0.66 \\
0.025 & 0.292(11) & 0.92 & 0.281(10) & 0.68      & 0.272(13) & 0.72 \\
0.03 & 0.314(8) & 0.93   & 0.306(9) & 0.88       & 0.298(11) & 0.80 \\
0.04 & 0.351(7) & 0.98   & 0.346(8) & 1.12       & 0.342(9) & 0.89 \\
0.05 & 0.384(6) & 1.06   & 0.381(7) & 1.23       & 0.379(8) & 0.97 \\
0.06 & 0.415(6) & 1.13   & 0.413(7) & 1.25       & 0.413(7) & 1.06 \\
0.08 & 0.473(6) & 1.27   & 0.473(6) & 1.20       & 0.476(6) & 1.24 \\
0.10 & 0.526(5) & 1.38   & 0.526(6) & 1.16       & 0.532(5) & 1.41 \\
0.13 & 0.601(4) & 1.52   & 0.600(5) & 1.16       & 0.609(5) & 1.69 \\
0.17 & 0.692(4) & 1.71   & 0.691(5) & 1.30       & 0.703(4) & 2.00 \\
0.23 & 0.819(3) & 2.00   & 0.817(4) & 1.76       & 0.832(4) & 2.16 \\ 
\hline
\end{tabular}  
\caption{{}\label{tab:m_pi_9x24} Pion masses for parametrized FP
  fermions on $9^3\times 24$ lattice from three different pion
  operators.}
\end{center}
\end{table}

\begin{figure}[htb]
\begin{center}
  \leavevmode \epsfxsize=120mm
  \epsfbox{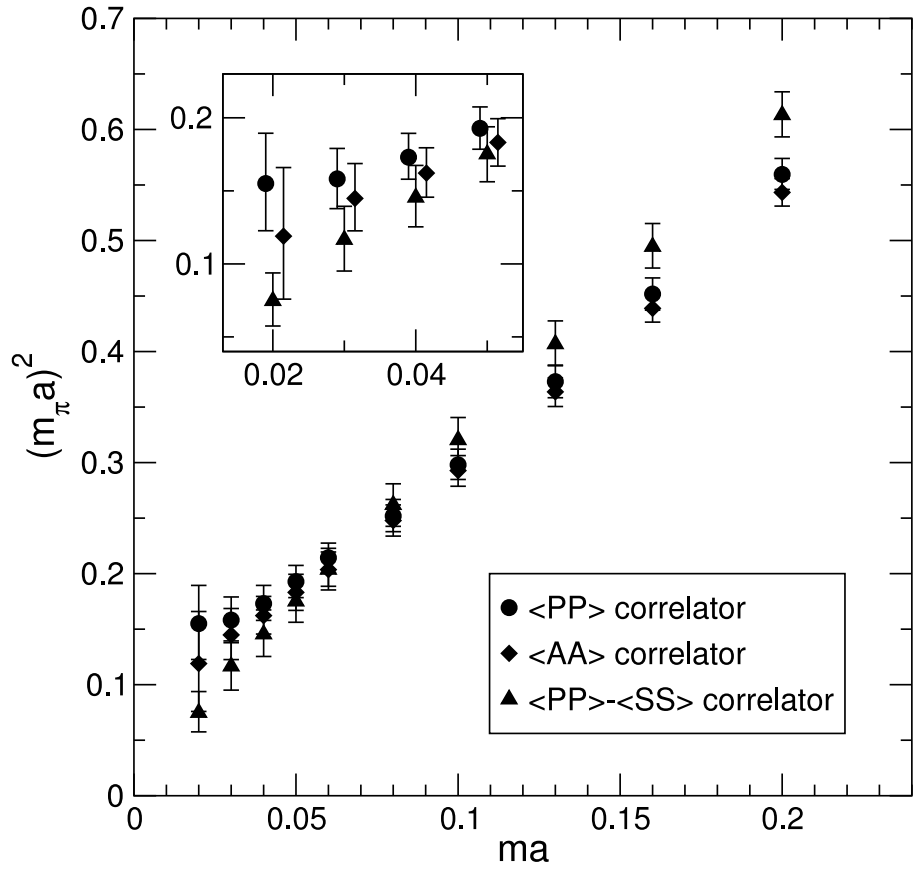}
  \vskip -6mm
\end{center}
\caption{{}\label{fig:fp_pisquared_6} Squared pion mass for 
  the parametrized FP Dirac operator $D^{\rm FP}$ from 100 
  $6^3\times 16$ gauge configurations.
  The inset shows the smallest four masses on a magnified scale to
  identify topological finite-volume effects.}
\end{figure}

\begin{figure}[htb]
\begin{center}
  \leavevmode \epsfxsize=120mm
  \epsfbox{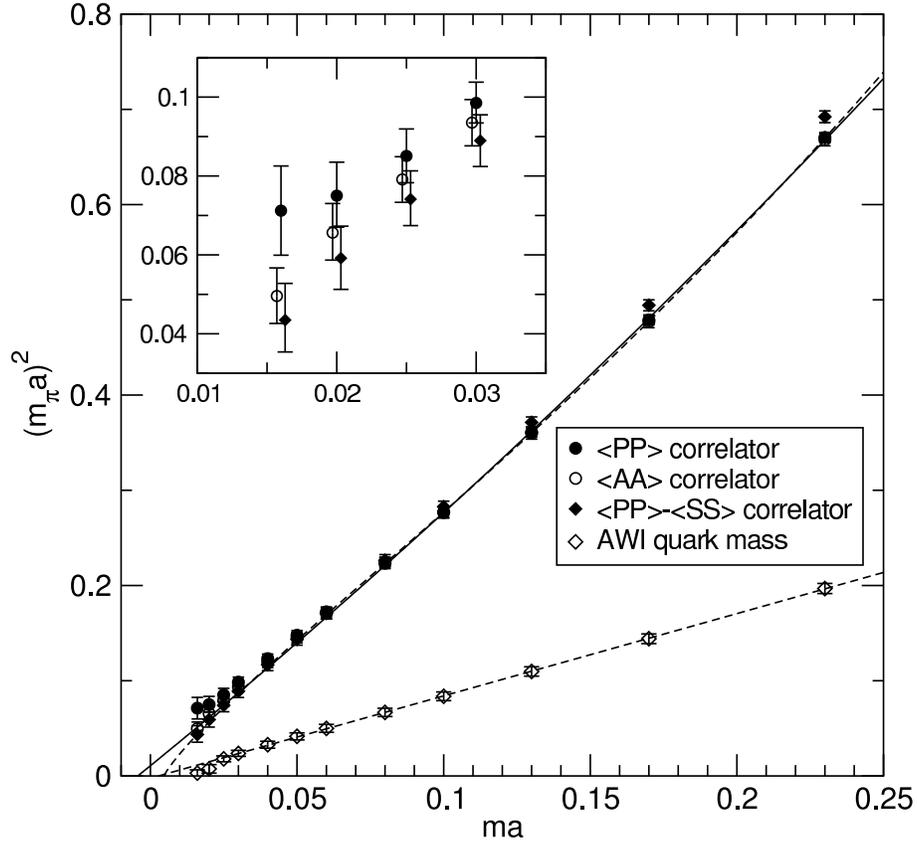}
  \vskip -6mm
\end{center}
\caption{{}\label{fig:fp_pisquared_9} Squared pion mass for 
  the parametrized FP Dirac operator $D^{\rm FP}$ from 70 
  $9^3\times 24$ gauge configurations.
  The massive Dirac operator is constructed with the covariant scalar
  density as in Eq.~\eqref{Act_mod}. Shown are quadratic fits to the P
  correlator at large and the P-S correlator at small quark mass with
  (dashed line) and without (solid line) the Q$\chi$PT logarithm. The
  inset shows the smallest four masses on a magnified scale.
  Topological finite-volume effects are reduced considerably. Also
  plotted is the unrenormalized quark mass from the axial Ward
  identity (AWI) together with a linear fit. When the quenched chiral
  logarithm is included in the fit, the fits from the AWI quark mass
  and from the squared pion mass agree in the chiral limit.}
\end{figure}

Figs.~\ref{fig:fp_pisquared_6} and \ref{fig:fp_pisquared_9} show a
comparison of the squared pion mass versus the input quark mass at the
two lattice sizes $6^3\times 16$ and $9^3\times 24$ using the
parametrized FP Dirac operator. On the smaller volume, the three
different pion correlators give very different values at small quark
masses. The P correlator lies highest, while the P-S correlator gives
considerably smaller pion masses, as the inset in
Fig.~\ref{fig:fp_pisquared_6} shows. The A correlator lies in between
the two. This behaviour is in qualitative agreement with the results
from domain-wall fermions \cite{Blum:2000kn}. For larger quark masses,
the P-S correlator deviates from the other two, as it is then
difficult to disentangle the contribution of the closely lying scalar
state in the mass determination, which can also be seen from the
increasing value of $\chi^2$ for the mass fit in 
Table~\ref{tab:m_pi_9x24}.
As expected, the discrepancy between the three
correlators in the chiral limit is reduced at our larger lattice
volume. The ordering among the correlators is still the same, but the
pion mass difference is smaller. A recent study with Wilson overlap
fermions at a similar spatial lattice size of $\approx 1.4\fm$ also
shows no significant difference \cite{Giusti:2001pk}, but their
smallest quark mass is much larger than ours. We can conclude from our
findings that at the rather small statistics that we have collected,
the topological finite-volume effects become non-negligible at the
smallest few quark masses, where the pseudoscalar correlator is
clearly contaminated and cannot be used to get reliable results. We
therefore decide to work in the following with the P-S correlator at
small and with the P correlator at large quark mass, changing
correlators at an intermediate quark mass where both agree, which is
at $ma=0.06$ for the $9^3 \times 24$ lattice.

\begin{table}[htb]
\begin{center}
\begin{tabular}{|l|l|lc|lc|} \hline

$ma$ & $m_\pi/m_\rho$& $m_\rho$ & $\chi^2/df$  & $m_N$ & $\chi^2/df$ \\ \hline
0.016 & 0.304(33) & 0.673(21) & 0.87     & 0.919(117)  &   0.66 \\
0.02 & 0.357(26) & 0.679(21) & 0.48      & 0.928(102) & 0.83 \\
0.025 & 0.396(20) & 0.686(14) & 1.06     & 0.908(74) & 0.95 \\
0.03 & 0.429(17) & 0.693(12) & 1.42      & 0.927(55) & 1.03 \\
0.04 & 0.482(15) & 0.708(12) & 1.79      & 0.975(43) & 1.43 \\
0.05 & 0.523(14) & 0.723(11) & 2.02      & 1.013(32) & 1.82 \\
0.06 & 0.562(11) & 0.735(10) & 2.20      & 1.043(27) & 2.38 \\
0.08 & 0.619(10) & 0.763(9) & 2.45       & 1.098(19) & 2.97 \\
0.10 & 0.666(9)  & 0.788(9) & 2.56       & 1.152(15) & 3.22 \\
0.13 & 0.726(8)  & 0.828(7) & 2.51       & 1.232(11) & 3.46 \\
0.17 & 0.786(7)  & 0.886(7) & 2.28       & 1.335(11) & 3.71 \\
0.23 & 0.852(6)  & 0.979(6) & 2.16      & 1.489(10) & 4.17 \\\hline
\end{tabular}
\end{center}
\caption{{}\label{tab:m_rho_9x24} $m_\pi/m_\rho$, rho and nucleon masses
   for parametrized FP fermions on $9^3\times 24$ lattice. }
\end{table}

On the larger lattice, we also measure the unnormalized AWI quark mass
\begin{equation} \label{eq:pcac_mass}
2m_q(t) = \frac{\sum_{\vec x} \langle\partial_4 A_4(\vec x,
 t)P(0)\rangle}{\sum_{\vec 
 x} \langle P(\vec x, t)P(0)\rangle},
\end{equation}
and take the average of the ratio of correlators over the range $5\leq
t \leq 9$. We use an ultralocal (non-conserved) axial current $A_\mu$
and neglect the renormalization factors $Z_A$ and $Z_P$ which would
show up on the right hand side of Eq.~\eqref{eq:pcac_mass}, as they
are not relevant for the following analysis. We fit the data with a
linear function to determine whether the remaining chiral symmetry
breaking of the action introduces a residual mass. The smallest two
masses were left out of the fit, which is shown as a dashed line in
Fig.~\ref{fig:fp_pisquared_9}. From the intersection of the linear fit
with the horizontal axis we read off a residual mass $m_{\rm res}^{\rm
  AWI}a=-0.003(4)$, which is consistent with zero within errors.  A
quadratic fit to the squared pion mass from the P-S correlator for
$am\leq 0.05$ and the P correlator for $am>0.05$ gives $m_{\rm
  res}a=0.004(4)$ with a value of $\chi^2/df=0.28$, while a fit to the
form predicted by quenched chiral perturbation theory (Q$\chi$PT)
\cite{Sharpe:1992ft}
\begin{equation}
m_{\rm PS}^2 = 2A(m_q+m_{\rm res}) \left[ 1 - \delta
\left( \ln\frac{2A(m_q+m_{\rm res})}{\Lambda_\chi^2} + 1 \right) \right] +
4B(m_q+m_{\rm res})^2,
\end{equation}
gives $m_{\rm res}a=-0.004(5)$ and $0.23(7)< \delta < 0.30(18)$ when
the scale $\Lambda_\chi$ is varied in the range between $0.8\GeV$ and
$1.2\GeV$, with typical values of $\chi^2/df=0.05$. Obviously the large
errors do not allow to single out a preferred form for the chiral fit,
but for the Q$\chi$PT form, the residual quark mass agrees with the
one from the axial Ward identity, while the agreement is worse in the
case of the quadratic fit.  This gives a hint that we see a signal of
the chiral logarithm in our pion mass measurements. (Note that the
smallness of the $\chi^2/df$ values is due to the fact that the
results for different quark masses are strongly correlated, hence an
improvement from $0.28$ down to $0.05$ is meaningful.)

\subsection{Energy-momentum dispersion relation}
\begin{figure}[htb]
\begin{center}
  \leavevmode \epsfxsize=100mm \epsfbox{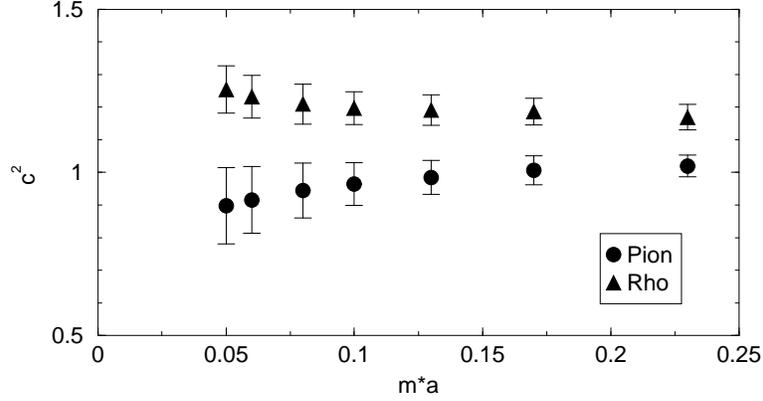}
  \vskip -6mm
\end{center}
\caption{{}\label{fig:c_squared} Squared speed of light $c^2$ 
  for pion and rho mesons with the parametrized FP Dirac operator 
  $D^{\rm FP} + m$ at the smallest non-zero momentum. 
  The error bars are from statistical errors only. 
  Systematic uncertainties from choosing the fit range
  (especially for the rho) would increase the errors considerably. }
\end{figure}

As the FP action is classically perfect, its dispersion relation is
expected to show small scaling violations. In Fig.~\ref{fig:c_squared}
the squared speed of light $c^2 = (E(p)^2 - m^2)/p^2$ measured on the
$9^3\times 24$ lattice is shown for the smallest momentum $\vec p = 2
\pi\vec n /9$ with $|\vec n| = 1$ ($|\vec p|\approx 0.8\GeV$) at
different quark masses. While $c^2$ for the pion is consistent with 1
within errors, it is not for the rho meson. However, the (purely
statistical) error bars do not include systematic uncertainties from
choosing the fit range, which was especially difficult for the rho in
this low statistics, small volume study.  As a qualitative result, it
is clear that compared to the Wilson or Sheikholeslami-Wohlert clover
action \cite{Lee:1997bq}, the dispersion relation is significantly
improved.

\begin{table}[htb]
\begin{center}
\begin{tabular}{|l|l|l|l|l|l|} \hline 

$ma$ & $m_\pi/m_\rho$& $m_\pi$(P)  & $m_\pi$(A)  & $m_\pi$(P-S) &
  $m_\rho$  \\ 
\hline
0.012 & 0.273(38) & 0.235(12) & 0.232(15) & 0.202(23) & 0.741(56) \\
0.016 & 0.313(32) & 0.257(19) & 0.258(13) & 0.234(20) & 0.747(43) \\
0.02 & 0.348(29) & 0.279(10) & 0.280(12) & 0.261(18) & 0.752(36) \\
0.03 & 0.418(24) & 0.326(8) & 0.328(11) & 0.319(14) & 0.763(27) \\
0.04 & 0.472(22) & 0.370(7) & 0.370(9) & 0.365(11) & 0.773(26) \\
0.06 & 0.556(18) & 0.444(6) & 0.443(7) & 0.443(8) & 0.797(22) \\
0.09 & 0.644(15) & 0.536(6) & 0.534(6) & 0.541(5) & 0.832(17) \\
0.13 & 0.723(12) & 0.640(5) & 0.639(6) & 0.651(5) & 0.886(13) \\
0.17 & 0.776(10) & 0.734(4) & 0.733(5) & 0.748(5) & 0.946(10) \\
0.23 & 0.829(7) & 0.863(4) & 0.862(5) & 0.880(4) & 1.042(8) \\ \hline

\end{tabular}
\end{center}
 \caption{{}\label{tab:ov_m_rho_9x24_3.0} $m_\pi/m_\rho$, 
   pion and rho masses for overlap-improved FP fermions on 
   $9^3\times 24$ lattice at $\beta=3.0$ 
   (lattice spacing $0.16\fm$). }
\end{table}

\section{First results for the overlap-improved FP action}
Using the parametrized FP Dirac operator $D^{\rm FP}$ as a starting
point for an overlap expansion, it is possible to remove any effects
from residual chiral symmetry breaking which is due to the
imperfection of the parametrization.  Note that the true FP Dirac
operator would reproduce itself by the overlap procedure. Since our
$D^{\rm FP}$ is close to the true FP Dirac operator we expect that the
overlap procedure will not drive it too much away from the FP, i.e. by
improving chiral properties we don't spoil the good scaling
properties.  We define the massive overlap Dirac operator (for a
discussion of the $2R \ne 1$ case, see Sect.~5)
\begin{equation}
  \label{eq:massive_overlap}
D_{\rm ov}(m) = (1-\frac{m}{2}) D_{\rm ov}(0) + \frac{m}{2R},
\end{equation}
where the massless overlap operator is
\begin{equation}
D_{\rm ov}(0) = \frac{1}{\sqrt{2R}}  
\left( 1- \frac{A}{\sqrt{A^\dagger A}} \right) \frac{1}{\sqrt{2R}},
\end{equation}
and in the kernel,
\begin{equation}
A = 1 - \sqrt{2R} D^{\rm FP} \sqrt{2R},
\end{equation}
the parametrized FP Dirac operator enters. $R$ is an explicit
parametrization of the fixed-point $R$ in the Ginsparg-Wilson relation
in Eq.~(\ref{gw}).

\begin{table}[htb]
\begin{center}
\begin{tabular}{|l|l|l|l|l|l|} \hline 

$ma$ & $m_\pi/m_\rho$& $m_\pi$(P)  & $m_\pi$(A)  & $m_\pi$(P-S) 
& $m_\rho$  \\ 
\hline
0.012 & 0.236(69) & 0.216(16) & 0.189(27) & 0.145(37) & 0.614(83) \\
0.016 & 0.297(61) & 0.232(14) & 0.214(22) & 0.176(30) & 0.592(66) \\
0.02 & 0.346(52) & 0.250(14) & 0.236(21) & 0.203(35) & 0.588(53) \\
0.03 & 0.446(41) & 0.296(11) & 0.287(19) & 0.264(18) & 0.592(36) \\
0.04 & 0.518(35) & 0.336(9) & 0.329(19) & 0.314(15) & 0.606(29) \\
0.06 & 0.624(28) & 0.406(9) & 0.405(16) & 0.398(11) & 0.639(23) \\
0.09 & 0.714(21) & 0.493(8) & 0.497(12) & 0.497(9) & 0.691(17) \\
0.13 & 0.782(17) & 0.593(7) & 0.596(8) & 0.605(7) & 0.758(14) \\
0.17 & 0.829(13) & 0.684(6) & 0.686(6) & 0.699(7) & 0.826(11) \\
0.23 & 0.873(10) & 0.811(6) & 0.811(5) & 0.828(6) & 0.929(9) \\ 
\hline
\end{tabular}
\end{center}
 \caption{{}\label{tab:ov_m_rho_9x24_3.2} $m_\pi/m_\rho$, 
   pion and rho masses for overlap-improved FP fermions on 
   $9^3\times 24$ lattice at $\beta=3.2$ 
   (lattice spacing $0.13\fm$). }
\end{table}

\subsection{Meson spectrum}
\begin{figure}[htb]
\begin{center}
  \leavevmode \epsfxsize=100mm
  \epsfbox{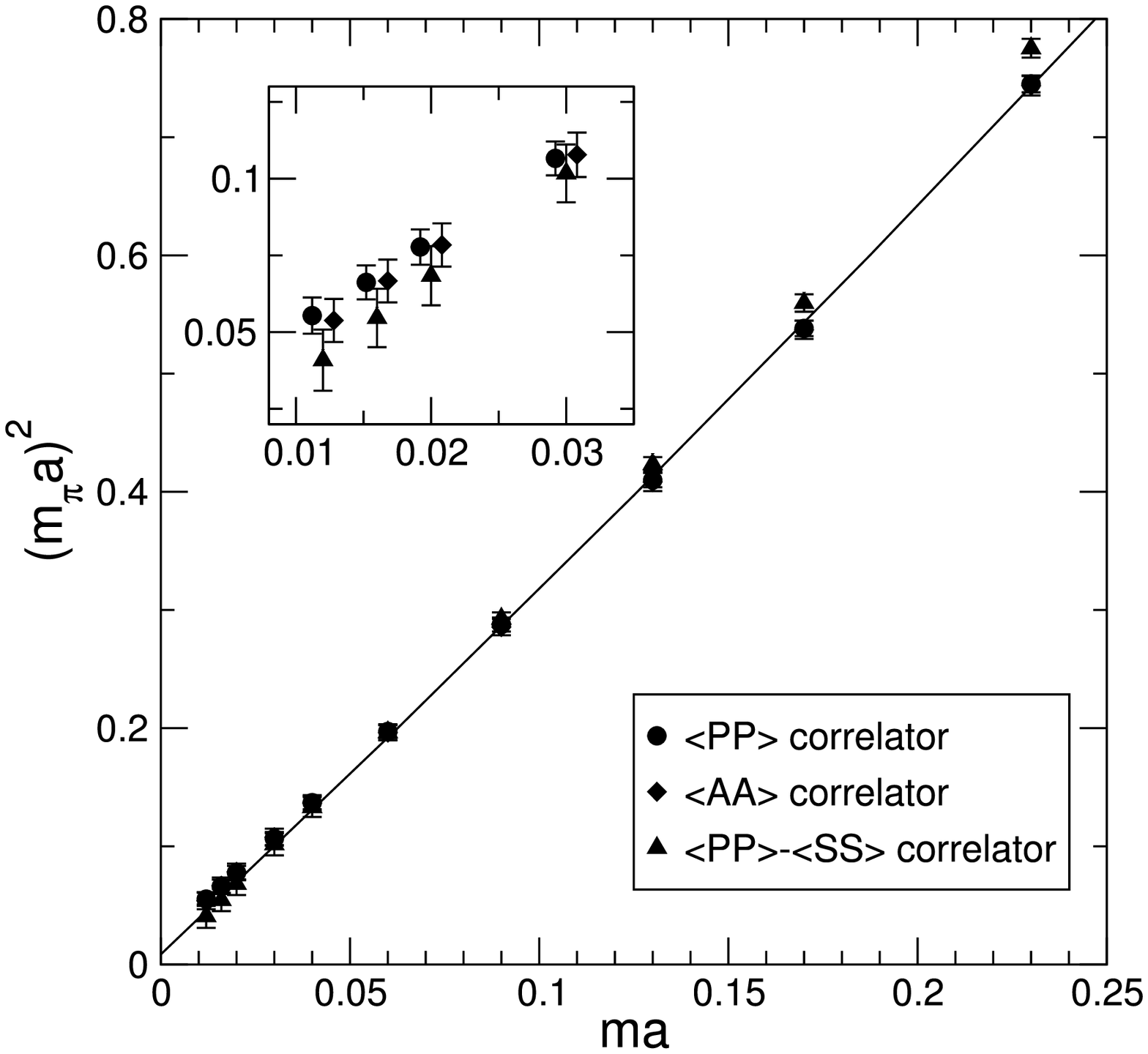}
  \vskip -6mm
\end{center}
\caption{{}\label{fig:sqrpion_overlap_b3.0} Squared pion mass for 
  the overlap-improved FP Dirac operator $D^{\rm FP}_{\rm ov}$ 
  from 28 $9^3\times 24$ gauge configurations at lattice spacing 
  $0.16\fm$, with a quadratic fit to the P correlator 
  at large and the P-S correlator at small quark mass. 
  The inset shows the smallest four masses on a magnified scale.}
\end{figure}

\begin{figure}[htb]
\begin{center}
  \leavevmode \epsfxsize=90mm \epsfbox{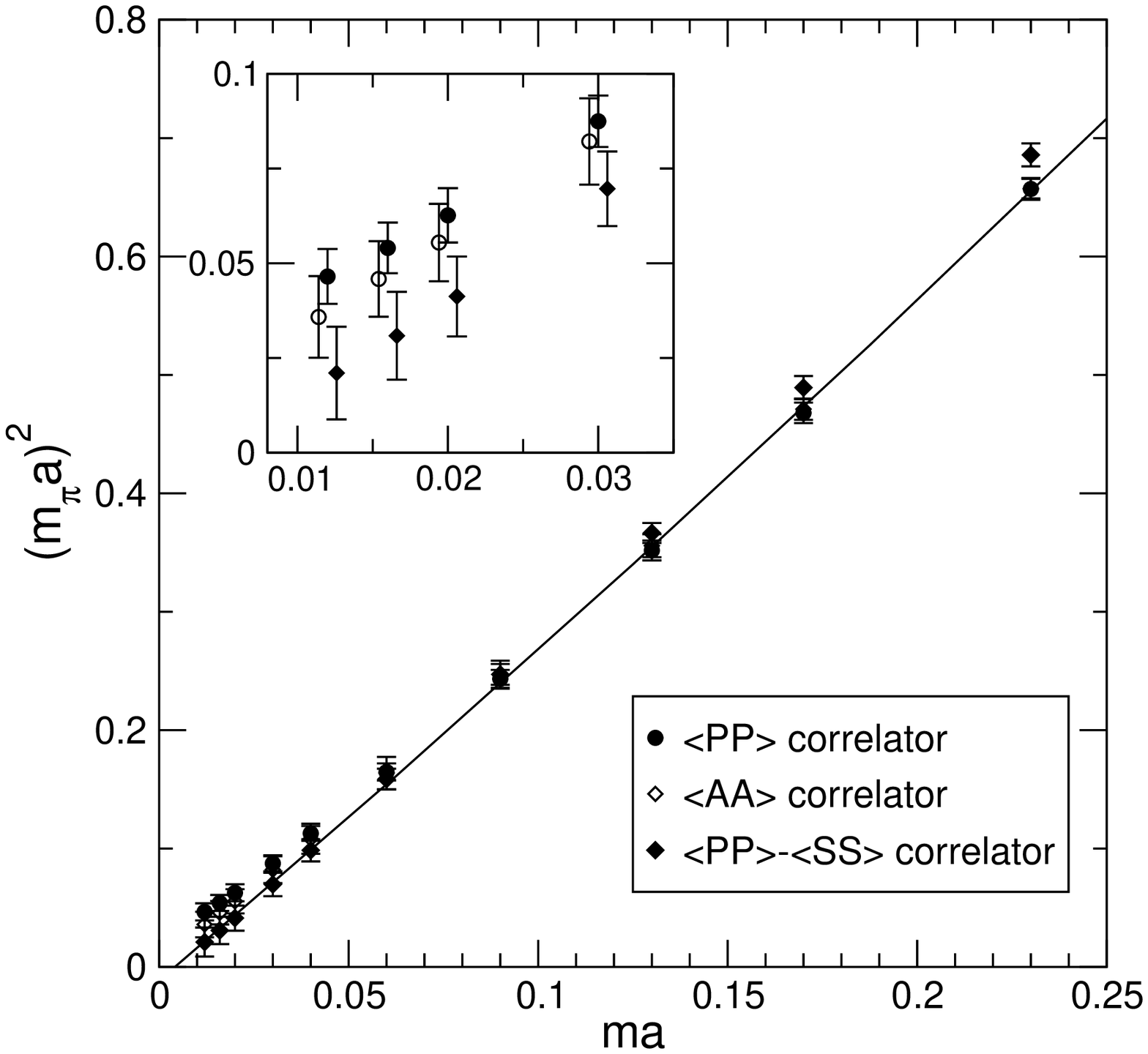}
  \vskip -6mm
\end{center}
\caption{{}\label{fig:sqrpion_overlap_b3.2} Squared pion mass 
  for the overlap-improved FP Dirac
  operator $D^{\rm FP}_{\rm ov}$ from 32 $9^3\times 24$ gauge
  configurations at lattice spacing $0.13\fm$, with a quadratic
  fit to the P correlator at large and the P-S correlator at small
  quark mass. The inset shows the smallest four masses on a magnified
  scale.}
\end{figure}

For spectroscopy we approximate the inverse square root
$1/\sqrt{A^{\dagger} A}$ with a polynomial of order 3, projecting out
the smallest 20 eigenvalues of $A^\dag A$ for faster convergence.  The
improved chiral behaviour allows the quark mass to be decreased even
further. The smallest input quark mass we considered is $ma=0.012$
where the pion to rho mass ratio is $m_\pi/m_\rho=0.27(4)$ and
$0.24(7)$ at lattice spacings $0.16$ and $0.13\fm$ respectively 
(Tables \ref{tab:ov_m_rho_9x24_3.0} and \ref{tab:ov_m_rho_9x24_3.2}).
In Figs.~\ref{fig:sqrpion_overlap_b3.0} and
\ref{fig:sqrpion_overlap_b3.2}, we show the pion mass squared measured
using the overlap-improved FP Dirac operator.  For an exactly chiral
action, the topological finite-volume effects cancel exactly for the
P-S correlator. A quadratic fit to the squared pion mass is consistent
with zero at $m=0$ within the large error at this small number of
gauge configurations.

\subsection{{}Chiral condensate ${\mathbf \langle \Psibar \Psi \rangle}$} 
It is the common expectation that
for QCD with a number $N_f \ge 2$ of massless quark flavours, the
chiral symmetry is spontaneously broken by a non-zero expectation
value for the chiral condensate $\langle \Psibar \Psi \rangle$.
Chiral perturbation theory ($\chi {\rm PT}$), which is based on this
assumption, is an excellent description of many low-energy QCD
phenomena \cite{Gasser:1983yg}. However, it is only possible via
lattice QCD to test from first principles if the symmetry is
spontaneously broken.

The leading order effective theory of $\chi {\rm PT}$ contains the low
energy constants $f_{\pi}$ and $\Sigma$.  In full QCD the
Gell-Mann-Oakes-Renner relation
\begin{equation}
\label{gmor}
f_{\pi}^2 m_{\pi}^2 = 4 m \Sigma,
\end{equation}
becomes exact in the $m \rightarrow 0$ chiral limit and the chiral
condensate $\langle \Psibar \Psi \rangle = \langle \bar{u} u
\rangle = \langle \bar{d} d \rangle = \dots$, defined at zero quark
mass, is equal to -$\Sigma$. The low energy constant $\Sigma$ depends
on the number of massless flavours $N_f$.

In quenched QCD ($N_f=0$), the relation $\langle \Psibar \Psi
\rangle = -\Sigma$ and Eq.~(\ref{gmor}) receive corrections even in
the chiral limit \cite{Sharpe:1992ft,Osborn:1998qb}.  Actually,
$\langle \Psibar \Psi \rangle$ and $m_{\pi}^2/m$ in the chiral
limit are not defined due to diverging quenched chiral logarithms.
This expectation seems to be confirmed in numerical studies of the
Banks-Casher relation \cite{Kiskis:2001zt}.

On the other hand, it is possible to study and determine the low
energy constant $\Sigma(N_f=0)$ in the quenched theory.  Under the
assumption that $\Sigma(N_f)$ is a smooth function of $N_f$ and
$\Sigma(N_f=0)$ is close to $\Sigma(N_f=3)$ (which is the standard
assumption when quenched results are used to estimate full QCD
quantities) we get an estimate for the chiral condensate $-\langle
\Psibar \Psi \rangle (N_f=3) = \Sigma(N_f=3) \approx
\Sigma(N_f=0)$.

One possibility to determine $\Sigma(N_f=0)$ is to study the chiral
condensate in a fixed topological sector with charge $Q$ in a finite
volume $V$ at finite quark mass $m$ \cite{Osborn:1998qb}. The volume
and the quark mass are chosen so that the finite size effects are
dominated by the pions with zero momentum. Using $\chi {\rm PT}$, or
random matrix theory, the fermion condensate at finite volume and
quark mass has been calculated in the continuum, both for full and
quenched QCD.  The quenched QCD condensate $\langle \Psibar \Psi
\rangle_{m,V,Q}$ is given by
\begin{equation}
- \langle \Psibar \Psi \rangle_{m,V,Q} 
= m V \Sigma^2 [I_{|Q|}(z) K_{|Q|}(z) + I_{|Q|+1}(z) K_{|Q|-1}(z)] + 
 \frac{|Q|}{m V}\, ,
\end{equation}
where $I_Q$ and $K_Q$ are modified Bessel functions, $z = m \Sigma V$
and the $N_f=0$, $V=\infty$, $m=0$ low energy constant $\Sigma$ is the
quantity we wish to measure.  By measuring $\langle \Psibar \Psi
\rangle_{m,V,Q}$ in different topological sectors at different masses
and volumes, the continuum prediction of the $m$ and $V$ dependence
can be used to extract $\Sigma$.

On the lattice we calculate the bare subtracted condensate by
measuring the trace
\begin{align}
  - \langle \Psibar \Psi \rangle_{m,V,Q}^{\rm sub} &= \frac{1}{V}
  \Big\langle {\rm Tr}' \Big[(D(m) 2R)^{-1} - 
  \frac{1}{2} D(0) D^{-1}(m) \Big] \Big\rangle \notag \\
  &= \frac{1}{V} \frac{1}{1-\frac{m}{2}} \Big\langle {\rm Tr}'
  \Big[(D(m) 2R)^{-1} - \frac{1}{2} \Big] \Big\rangle \, ,
\label{eq:condensate}
\end{align}
where $D$ and $R$ are related via the Ginsparg-Wilson relation and
where we make use of Eq.~\eqref{eq:massive_overlap} to simplify the
expression on the first line.  Notice that in order to keep the
conventional notation we use  $\langle \Psibar \Psi
\rangle_{m,V,Q}^{\rm sub}$ in Eq.~\eqref{eq:condensate} even though we
actually measure the expectation value of the scalar density $S^{0}$
as defined in Eq.~\eqref{SP}, which differs from $\langle \Psibar
\Psi \rangle_{m,V,Q}^{\rm sub}$ by a tree level factor that will
finally drop out in the renormalized value for $\Sigma$.  
We measure the trace stochastically using random $Z(2)$ 
vectors \cite{Dong:1993pk}.  In order to measure the condensate at 
very small quark mass, the remnant explicit chiral symmetry breaking 
must be very small, so we use the overlap-improved operator 
$D_{\rm ov}^{\rm FP}$, where $D^{\rm FP}$ is taken as input. 
Due to the $|Q|$ chiral zero modes of the Dirac operator, 
the quenched condensate contains a term $|Q|/(mV)$ which
diverges as the mass becomes small. In the volumes we study, the $|Q|$
zero modes always have the same chirality and their contribution to
the condensate is removed by measuring the trace ${\rm Tr}'$ in the
chiral sector opposite to the zero modes
\cite{Edwards:1998hh,Hernandez:1999gg}, i.e.~if the $|Q|$ modes have
chirality $+$, the $Z(2)$ vectors used to measure the trace are chosen
to have chirality $-$.  To determine $\langle \Psibar \Psi
\rangle_{m,V,Q}^{\rm sub}$, the stochastic trace is doubled to include both
chiral sectors.

We first generated ensembles of gauge configurations at different
volumes.  We determined the topological charge $Q$ of the
configurations from the chirality of the lowest eigenmodes of 
$(D^{\rm FP}_{\rm ov})^{\dagger} D^{\rm FP}_{\rm ov}$, which we find 
with an Arnoldi solver \cite{Sorenson:1992}.  To scan for the topology,
a Legendre polynomial of order 2 is used to approximate
$1/\sqrt{A^{\dagger}A}$, projecting out the 10 smallest $A^{\dagger}A$
eigenvalues. This is sufficiently accurate to determine the chirality
of the eigenmodes to better than $1\%$ accuracy.

\begin{table}[thb]
\begin{center}
\begin{tabular}{|c|c|c|c|} \hline 

$L$ & order & $|Q|$ & $N_{\rm conf}$ \\ \hline \hline

8  & 7  & 1 & 154 \\
   &    & 2 & 41  \\ \hline
10 & 10 & 1 & 53  \\
   &    & 2 & 43  \\
\hline

\end{tabular}
\end{center}
 \caption{{}\label{tab:condensate_meas} Polynomial order of 
   $1/\sqrt{A^{\dagger} A}$ approximation and statistics for 
   the condensate measurement.}
\end{table}

\begin{figure}[htb]
\begin{center}
  \vskip -1mm \leavevmode \epsfxsize=90mm \epsfbox{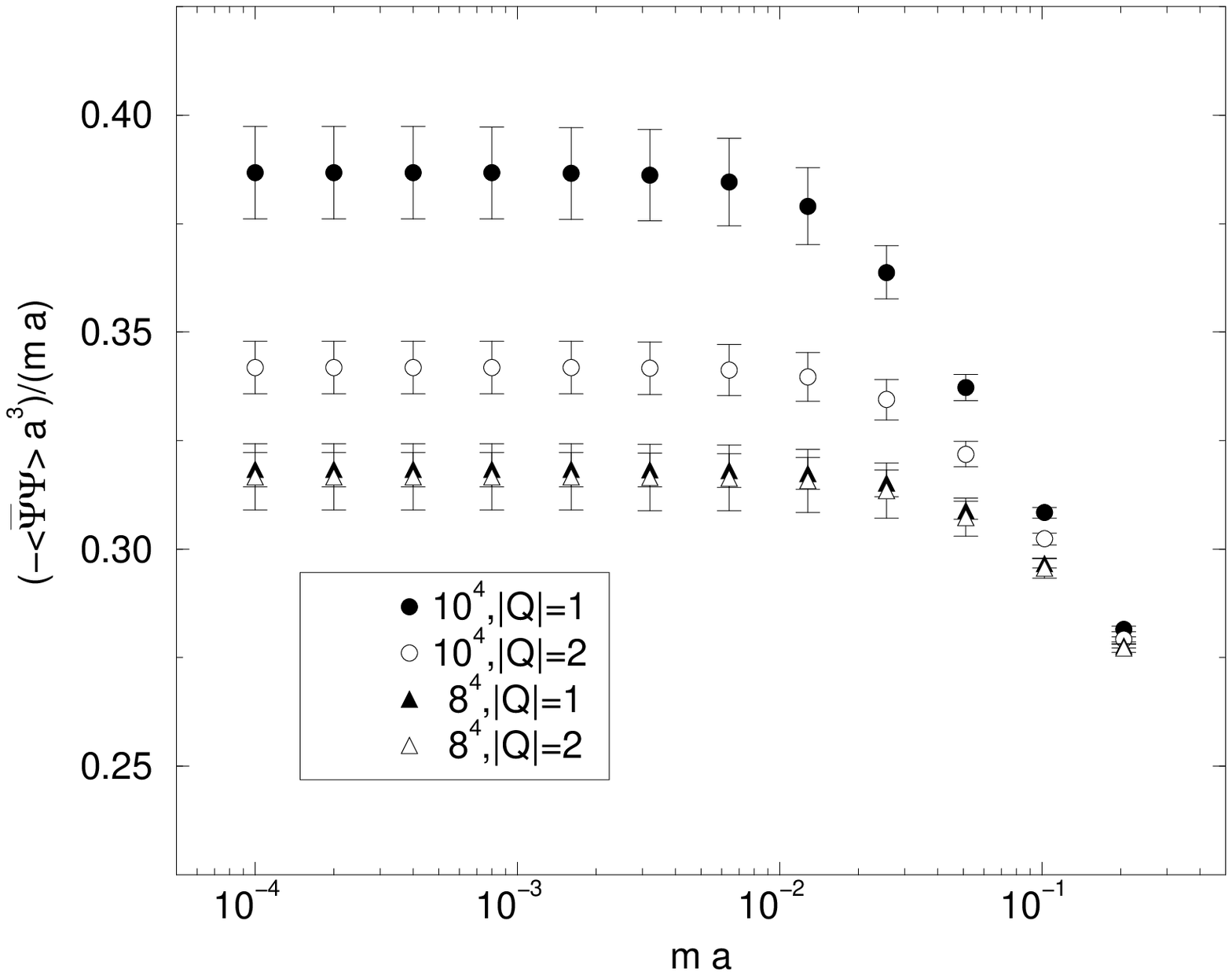}
  \vskip -6mm
\end{center}
\caption{{}\label{fig:condensate_m} The ratio 
  $(-\langle \Psibar \Psi \rangle_{m,V,Q}^{\rm sub} a^3)/(m a)$ 
  versus $m a$ for different volumes and topological sectors. }
\end{figure}

We have measured the condensate in volumes $8^4$ and $10^4$ at lattice
spacing $0.13\fm$ (corresponding to the bare coupling
$\beta=3.2$). We use 10 random $Z(2)$ vectors to measure the trace for
each configuration and a BiCGstab multi-mass solver to invert $D_{\rm
  ov}^{\rm FP}$ at all masses simultaneously. In the overlap operator,
we approximate $1/\sqrt{A^{\dagger}A}$ with Legendre polynomials of
order 7 and 10 for volumes $8^4$ and $10^4$, respectively. This gives
sufficiently precise chiral symmetry --- increasing the polynomial
order further, the relative change in $(\langle \Psibar \Psi
\rangle_{m,V,Q}^{\rm sub} a^3)/(m a)$ is $\leq {\cal O}(10^{-4})$.  
We project out the 10 smallest $A^{\dagger}A$ eigenvalues, which are 
treated exactly. Our statistics are given 
in Table~\ref{tab:condensate_meas}.
In Fig.~\ref{fig:condensate_m}, we plot $(-\langle \Psibar \Psi
\rangle_{m,V,Q}^{\rm sub} a^3)/(m a)$ as a function of $m a$ for 
the different volumes and topological sectors.

\begin{figure}[htb]
\begin{center}
  \vskip -3mm \leavevmode \epsfxsize=100mm \epsfbox{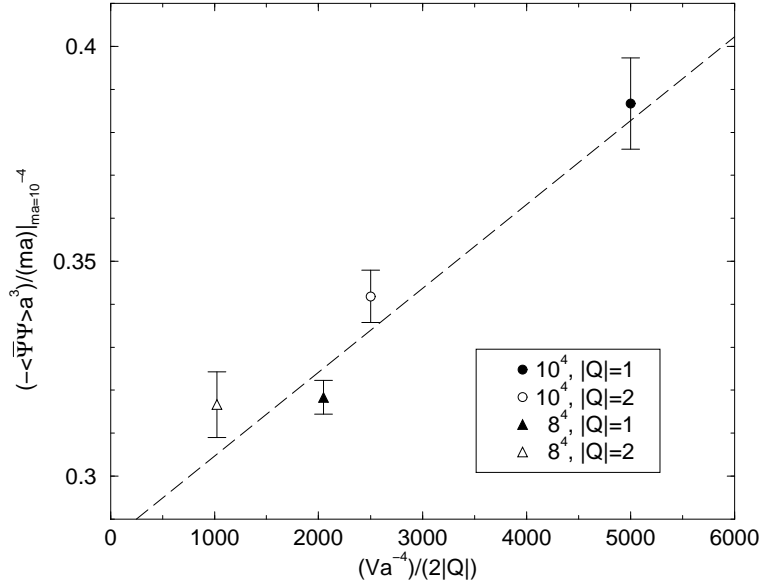}
  \vskip -6mm
\end{center}
\caption{{}\label{fig:condensate_V} The ratio 
  $(-\langle \Psibar \Psi \rangle_{m,V,Q}^{\rm sub} a^3)/(m a)$ 
  at $m a=10^{-4}$ versus $V/(2|Q|)$. The dashed line is a $\chi^2$-fit.}
\end{figure}

The bare quark condensate at finite quark mass contains a $\sim m/a^2$
cut-off effect. As the quark mass $m \rightarrow 0$,
\begin{equation}
- \frac{\langle \Psibar \Psi \rangle_{m,V,Q}^{\rm sub}}{m} = 
 \frac{\Sigma^2 V}{2 |Q|} + \frac{c_1}{a^2} ,
\label{eq:condensate_fit}
\end{equation}
where $c_1$ is an unknown coefficient which has to be fitted.  As the
coefficient $c_1$ comes from ultraviolet fluctuations, it is natural
to assume that it is independent of the topological charge $Q$.  The
contribution of the $|Q|$ zero modes has been removed, and there is no
artifact $1/(m a^3)$ in Eq.~\eqref{eq:condensate_fit} due to the fact
that the condensate is defined as the expectation value of a scalar
operator which transforms covariantly under chiral transformations and
has no mixing with the unit operator. (For further discussions, see
Sec.~5.) From Fig.~\ref{fig:condensate_m}, we see that $(-\langle
\Psibar \Psi \rangle_{m,V,Q}^{\rm sub} a^3)/(m a)$ reaches a 
plateau at very small quark mass. 
In Fig.~\ref{fig:condensate_V}, we plot the value of
$(-\langle \Psibar \Psi \rangle_{m,V,Q}^{\rm sub} a^3)/(m a)$ 
at $m a=10^{-4}$ versus $V/(2|Q|)$, which we fit to the form of
Eq.~\eqref{eq:condensate_fit}.  From the slope, we extract the bare
low energy constant as $a^3 \Sigma=4.42(36) \times 10^{-3}$.  We
convert this using the Sommer parameter ($r_0/a=3.943(60)$ has
previously been measured at this lattice spacing), giving $r_0^3
\Sigma=0.271(22)(12)$, where the first error is statistical and the
second the uncertainty in the scale.

In order to turn this bare result into the renormalized low energy
constant we need the scalar renormalization factor $Z_S$.  In the
present test study we obtained $Z_S$ combining the continuum
extrapolated renormalization group invariant (RGI) quark mass of the
ALPHA collaboration \cite{Garden:1999fg} with our spectroscopy data in
Sect.~4.1. This method has been suggested recently by Hern\'andez et
al.~\cite{Hernandez:2001yn}.

Chiral symmetry connects the scalar and pseudoscalar renormalization
factors: $Z_{P} = Z_S = 1/Z_m$, where $Z_m$ is the multiplicative mass
renormalization. This way the problem is reduced to finding $Z_m$.
The technique requires the measurement of the pion mass at a number of
quark masses. The renormalization factor $Z_m$ connecting our quark
mass $m$ to the RGI mass M defined at some reference pion mass
$m_{\pi}$ is given by
\begin{equation}
  Z_m = \left.\frac{U_M}{r_0 m}\right|_{(r_0 m_{\pi})^2=x_{\rm ref}},
\end{equation}
where $U_M=r_0 M$. The renormalization factor $Z_m$ should be
independent of the reference point, so finding this ratio at a number
of pseudoscalar masses indicates the systematic error. We have
performed mass measurements at two lattice spacings corresponding to
$\beta=3.0$ and $3.2$.  We use the same reference points and values of
$U_M$ as in \cite{Hernandez:2001yn} to determine $Z_m$. The results
are summarized below in Table~\ref{tab:renorm}.

\begin{table}[htb]
\begin{center}
\begin{tabular}{|l|l|l|l|l|} \hline

$\beta$ & $x_{\rm ref}$ & $U_M$ & $r_0 m$ & $Z_m$ \\ \hline \hline
3.0 & 1.5736 & 0.181(6)  & 0.156(4) &  1.16(5) \\
    & 3.0    & 0.349(9)  & 0.296(8) &  1.18(5) \\
    & 5.0    & 0.580(12) & 0.490(14) & 1.18(4) \\ \hline
3.2 & 1.5736 & 0.181(6)  & 0.141(27) & 1.28(25) \\
    & 3.0    & 0.349(9)  & 0.271(24) & 1.29(12) \\
    & 5.0    & 0.580(12) & 0.446(22) & 1.30(7) \\ \hline
 
\end{tabular}
\end{center}
 \caption{{}\label{tab:renorm}The renormalization factor $Z_m$ 
   determined at different 
   lattice spacings and reference pseudoscalar masses. }
\end{table}

At each lattice spacing, the values for $Z_m$ at the different
reference points are in very good agreement with one another. We take
the average of the values and the error at $x_{\rm ref}=3.0$ as our
determination of $Z_m$.  The renormalization group invariant low
energy constant is $\hat{\Sigma}= Z_S \Sigma$. To convert this result
into the $\overline{\rm MS}$ scheme, we use the fact that
\begin{equation}
\frac{\hat{\Sigma}}{\Sigma_{\overline{\rm MS}}(2\GeV)} 
= \frac{\overline{m}_{\overline{\rm MS}}(2\GeV)}{M}
=0.72076,
\end{equation}
where the ratio of masses has been calculated perturbatively to four
loops in the $\overline{\rm MS}$ scheme \cite{Capitani:1998mq}.

\begin{figure}[htb]
\begin{center}
  \vskip -3mm \leavevmode \epsfxsize=100mm \epsfbox{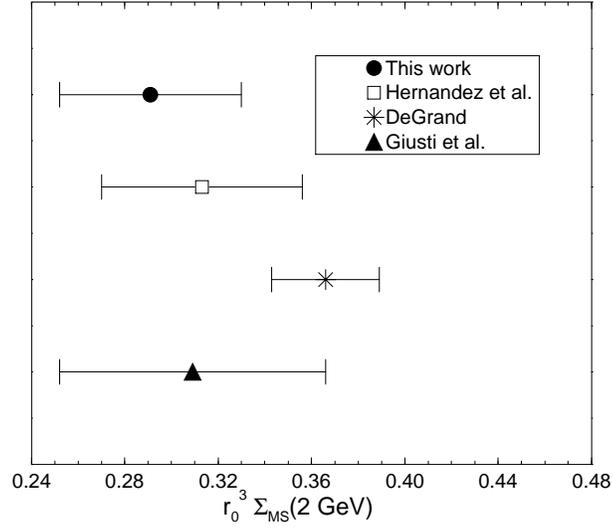}
  \vskip -6mm
\end{center}
\caption{{}\label{fig:condensate_MSbar} Comparison of different 
  measurements of $r_0^3 \Sigma_{\overline{\rm MS}}(2\GeV)$.}
\end{figure}

Taking the renormalization factor $Z_S=1/Z_m=0.78(7)$ at $\beta=3.2$,
where the bare $\Sigma$ is obtained, we get for the renormalization
group invariant low energy constant $r_0^3 \hat{\Sigma} =
0.210(17)(9)(20)$, with statistical, scale and renormalization errors
respectively.  Taking $r_0=0.5\fm$ and combining the errors in
quadrature, this corresponds to 
$\hat{\Sigma}=(235 \pm 11\MeV)^3$.  
We can also convert this result to the $\overline{\rm MS}$ 
scheme giving 
$\Sigma_{\overline{\rm MS}}(2\GeV)= (262 \pm 12\MeV)^3$. 
As Fig.~\ref{fig:condensate_MSbar} illustrates,
our measurement of the renormalized chiral condensate is in good
agreement with other recent determinations
\cite{Giusti:2001pk,Hernandez:1999gg,DeGrand:2001ie}.

\subsection{Alternative determinations of the chiral condensate}
A direct determination of $\Sigma$ discussed above is only possible
with a chirally symmetric action. An alternative, more
phenomenological way is to use the GMOR relation, Eq.~\eqref{gmor}
using the pion mass measurements and the experimental value of
$f_\pi$ \cite{Bochicchio1985,Giusti1999,Giusti:2001pk}.
Although the ratio $m_\pi^2/m_q$ is not defined in
the quenched theory in the chiral limit $m_q \rightarrow 0$, due to
quenched chiral logarithms, we see that the pion mass squared is
consistent with linear behavior for intermediate quark mass.  Assuming
that the chiral logarithms have very little effect in this mass range,
we identify the slope $B_M$ as $4 \Sigma/f_{\pi}^2$. 
Table~\ref{tab:sigma_pion} gives the value of the slope $B_M$
and the bare condensate $a^3\Sigma$ calculated through the GMOR
relation, $\Sigma=B_M f_\pi^2/4$, using the physical value 
$f_\pi=131\MeV$. The value at $\beta=3.2$, 
$a^3 \Sigma=5.8(2)\times 10^{-3}$ has to be compared with the direct 
measurement discussed above, which gave $4.4(4)\times 10^{-3}$.

\begin{table}[thb]
\begin{center}
\begin{tabular}{|c|c|c|c|} 
\hline
$\beta$ & $a$[fm] & $B_M$    & $a^3 \Sigma\times 10^3$ \\  \hline \hline
3.0     & 0.16    & 3.22(12) & 8.8(3)            \\  \hline
3.2     & 0.13    & 2.90(12) & 5.8(2)            \\  \hline
 \end{tabular}
\end{center}
 \caption{{}\label{tab:sigma_pion} The value of the bare chiral 
   condensate determined from pion mass measurements. 
   The slope $B_M$ of $m_\pi^2$ vs. $m_q$, and the value of the bare 
   chiral condensate determined from GMOR relation, with
   $f_\pi=131\MeV$.
}
\end{table}

We have to mention a difficulty with the technique used in the direct
determination of the condensate discussed in the previous subsection.
Finding the finite-volume behavior of the trace in 
Eq.~\eqref{eq:condensate}, the major problem is from rare,
but very large contributions to the average.  These come from
configurations where the Dirac operator has very small, but non-zero,
eigenvalues. Due to their presence, we found it difficult to control 
the statistical error. The distribution of the low-lying eigenvalues of
the Dirac operator offers an alternative method to determine $\Sigma$
\cite{Nishigaki:1998is}. This method does not suffer from the problem
mentioned above. The appearance of very small eigenvalues does not
bring any large contributions, which might make this method
competitive, or even better than the finite-volume technique.

\begin{figure}[htb]
\begin{center}
  \vskip -3mm \leavevmode \epsfxsize=100mm \epsfbox{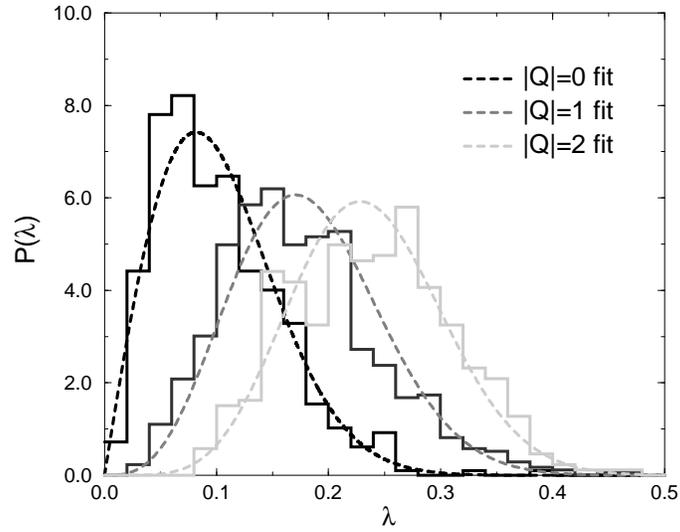}
  \vskip -6mm
\end{center}
\caption{{}\label{fig:RMT_b240} Distributions of the smallest 
  eigenvalue $\lambda$ of $D^{\rm FP}_{\rm ov}$ for topological 
  sectors $|Q|=0,1,2$, volumes $4^4$ at lattice spacing 
  $0.30\fm$. The curves are the random
  matrix theory fit to the histograms. }
\end{figure}

\begin{figure}[htb]
\begin{center}
  \vskip -3mm \leavevmode \epsfxsize=100mm \epsfbox{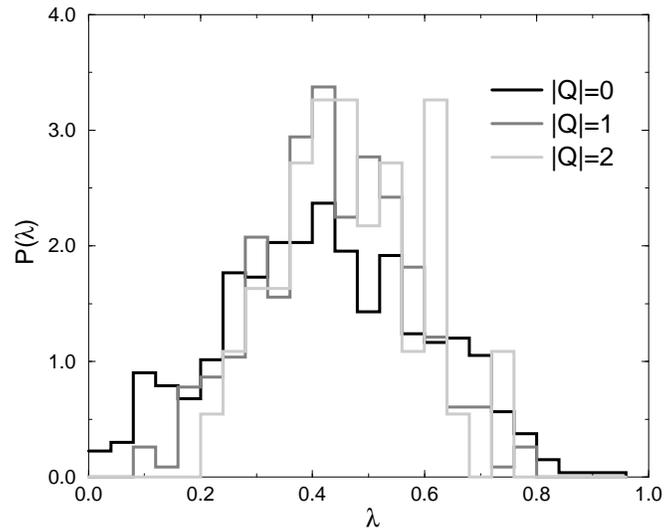}
  \vskip -6mm
\end{center}
\caption{{}\label{fig:RMT_b270} Distributions as 
  in Fig.~\ref{fig:RMT_b240} for volumes $4^4$ at lattice spacing 
  $0.22\fm$.}
\end{figure}

To compare techniques of measuring the low energy constant $\Sigma$,
we performed a test study of the distribution of low-lying eigenvalues
of the Dirac operator. We generated ensembles of $4^4$ lattice volume,
with 2000 gauge configurations at $\beta=2.4$ 
(lattice spacing $0.30\fm$) and 1000 configurations at $\beta=2.7$ 
(lattice spacing $0.22\fm$).
Admittedly, this lattice volume is too small and the resolutions 
are too coarse to allow a serious study of the low lying eigenvalues.
Using the Ritz functional method, we measured the smallest non-zero
eigenvalue of $D_{\rm ov}^{\rm FP}$ and determined the smallest
eigenvalue distribution for each topological sector. For the
overlap-improved operator, we used an order 5 polynomial approximation
of $1/\sqrt{A^\dagger A}$, with the smallest 20 and 8 $A^\dagger A$
eigenvalues projected out at $\beta=2.4$ and 2.7 respectively.  Even
on these coarse lattices, this low-order approximation has very little
chiral symmetry breaking. Using random matrix theory, the distribution
of the smallest eigenvalue $\lambda$ in topological sector $Q$ in the
quenched theory is
\begin{eqnarray} \label{eq:rmt}
P(z) &=& \frac{z}{2} e^{-z^2/4}~{\rm det}[ I_{2+i-j}(z) ] 
\hspace{1cm} i,j = 1,...,|Q| \nonumber \\
P(z) &=& \frac{z}{2} e^{-z^2/4} \hspace{3.2cm} Q=0, 
\end{eqnarray}
where $z=\lambda \Sigma V$ is the rescaled eigenvalue and $I_k$ are
the modified Bessel functions.  In Fig.~\ref{fig:RMT_b240}, we plot
the measured distributions for topological sectors $|Q|=0,1,2$ at
$\beta=2.4$, as well as the fits to the predicted form. We see the
fits describe the data quite well, allowing us to estimate the bare
quantity $a^3 \Sigma = 0.0687(27)$, or alternatively $r_0^3 \Sigma =
0.332(16)$. 
In Fig.~\ref{fig:RMT_b270}, however, we see that the smallest eigenvalue
distributions at $\beta=2.7$ for different topological sectors lie on
top of one another and are clearly not described by
Eq.~\eqref{eq:rmt}.  The extreme values of $\beta$ and/or lattice size
($L=4$) might be the reason for this result. Further studies are
needed to clarify this point.

\subsection{The quenched topological susceptibility}
\begin{figure}[htb]
\begin{center}
  \vskip -3mm \leavevmode \epsfxsize=120mm
  \epsfbox{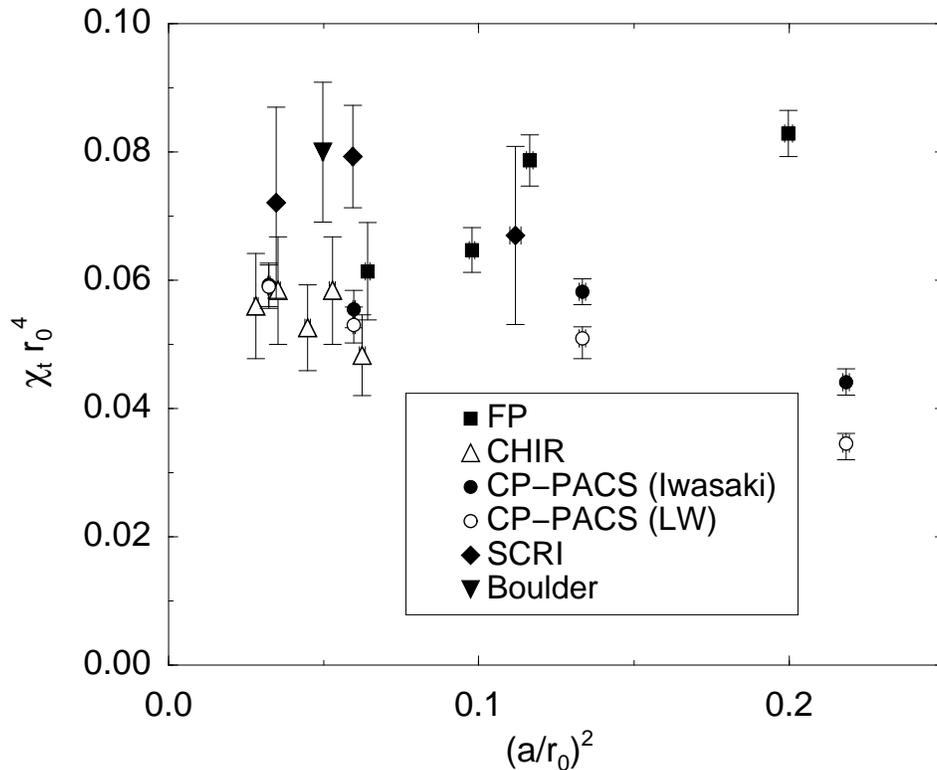} \vskip -6mm
\end{center}
\caption{{}\label{fig:top_susc} The quenched topological 
  susceptibility measured by different actions and techniques. 
  The CP-PACS data are obtained by cooling using the Iwasaki and 
  the L\"uscher-Weisz actions (see \cite{AliKhan:2001ym}). 
  The other data are obtained by chirally symmetric actions.
  FP denotes the present determination using $D_{\rm ov}^{\rm FP}$.
  The others are obtained using a chirally improved action
  \cite{Gattringer:2002mr}(CHIR), the overlap with the Wilson action
  \cite{Edwards:1998vu}(SCRI) and the overlap with the planar action
  \cite{DeGrand2000,DeGrand2002}(Boulder). }
\end{figure}

As discussed above, in the course of measuring the quark condensate 
we have determined the topological charge of the gauge configurations 
from the chirality of the lowest eigenmodes of 
$(D^{\rm FP}_{\rm ov})^{\dagger} D^{\rm FP}_{\rm ov}$.
As a byproduct, we obtain the quenched topological susceptibility 
$\chi_t=\langle Q^2 \rangle/V$ from the distribution of 
the topological charge. 
From an ensemble of 200 $\,10^4$ configurations at our smallest 
lattice spacing $0.13\fm$, we find $r_0^4 \chi_t=0.0612(75)$, 
corresponding to $\chi_t=(196 \pm 6\MeV)^4$. 
This result, which still might have sizeable cut-off effects, 
is consistent with earlier determinations, as shown in
fig.~\ref{fig:top_susc}.

\subsection{Local chirality of near-zero modes}
Exact zero modes of the Dirac operator tell us, via the index theorem,
about the topological charge $Q$ of the background gauge
configuration.  However, the exact zero modes alone cannot break
chiral symmetry spontaneously. According to the Banks-Casher relation
\cite{Banks:1979yr} $\langle \Psibar \Psi \rangle = - \pi \rho(0)
\ne 0$, the Dirac operator must build up a finite density of near-zero
modes, which does not vanish as $V \rightarrow \infty$.

One mechanism which explains the formation of near-zero modes involves
instantons. Consider a gauge configuration containing one instanton
and one anti-instanton. If the instanton and anti-instanton are
separated by a large distance, the Dirac operator has a pair of
complex eigenvalues lying close to 0. The farther the instanton and
anti-instanton are separated from each other, the closer the complex
eigenvalue pair moves to the origin. If the instanton and
anti-instanton are brought closer together, the complex eigenvalue
pair moves away from the origin and disappears into the bulk of the
eigenvalue spectrum. If the gauge configurations contain many
instantons and anti-instantons, this could produce a non-zero density
of near-zero modes, giving $\rho(0) \ne 0$ in the infinite volume
limit.

The question has recently been raised if it is possible to show that
the near-zero modes are dominated by instantons. From instanton
physics, it is expected that the modes are highly localized where the
instantons and anti-instantons sit. If this is so, then in these
regions the modes should be close to chiral i.e.~mostly either left-
or right-handed, depending on whether it is sitting on an instanton or
anti-instanton. In \cite{Horvath:2001ir}, the authors defined a
measure of local chirality at lattice site $x$ by
\begin{equation}
\tan\left[\frac{\pi}{4}(1+X(x))\right] = \sqrt{\frac{{\Psi_{\rm L}}^{\dagger} 
     \Psi_{\rm L}(x)}{{\Psi_{\rm R}}^{\dagger} \Psi_{\rm R}(x)}},
\end{equation}
where $\Psi_{\rm L/R}(x)$ are the standard ${\rm L/R}$ projections of
the corresponding wave function.  An exact zero mode is purely either
left- or right-handed, giving $X(x)=\pm 1$ at all lattice sites $x$.
If a near-zero mode is localized around instanton-anti-instanton
lumps, then $X(x)$ should be close to $\pm 1$ for the sites $x$ where
the probability density $\Psi^{\dagger} \Psi (x)$ is largest.

In the original paper \cite{Horvath:2001ir}, many near-zero modes of
the Wilson operator $D^{\rm W}$ were analysed for many gauge
configurations and the finding was that in the regions where the modes
are localized, the distribution for $X(x)$ is peaked around 0 and the
modes do not display local chirality. This led to the conclusion that
the near-zero modes are not dominated by instantons.  Since then,
several other groups have found the opposite conclusion
\cite{DeGrand:2001pj}, using a Dirac operator with much better chiral
symmetry than $D^{\rm W}$ (or even just an alternative definition of a
complete basis for the non-normal operator $D^{\rm W}$). They find the
distribution of $X(x)$ is double-peaked with maxima at large positive
and negative values of $X$, indicating that the modes are locally
chiral. On the other hand, in several more detailed comparisons the
semiclassical expectations were not confirmed by the numerical
data\cite{Horvath}. Presumably, one must not take this picture too
seriously in a situation where these objects do not form a dilute gas
and live in a strongly fluctuating background.

\begin{figure}[htb]
\begin{center}
  \vskip -3mm \leavevmode \epsfxsize=120mm
  \epsfbox{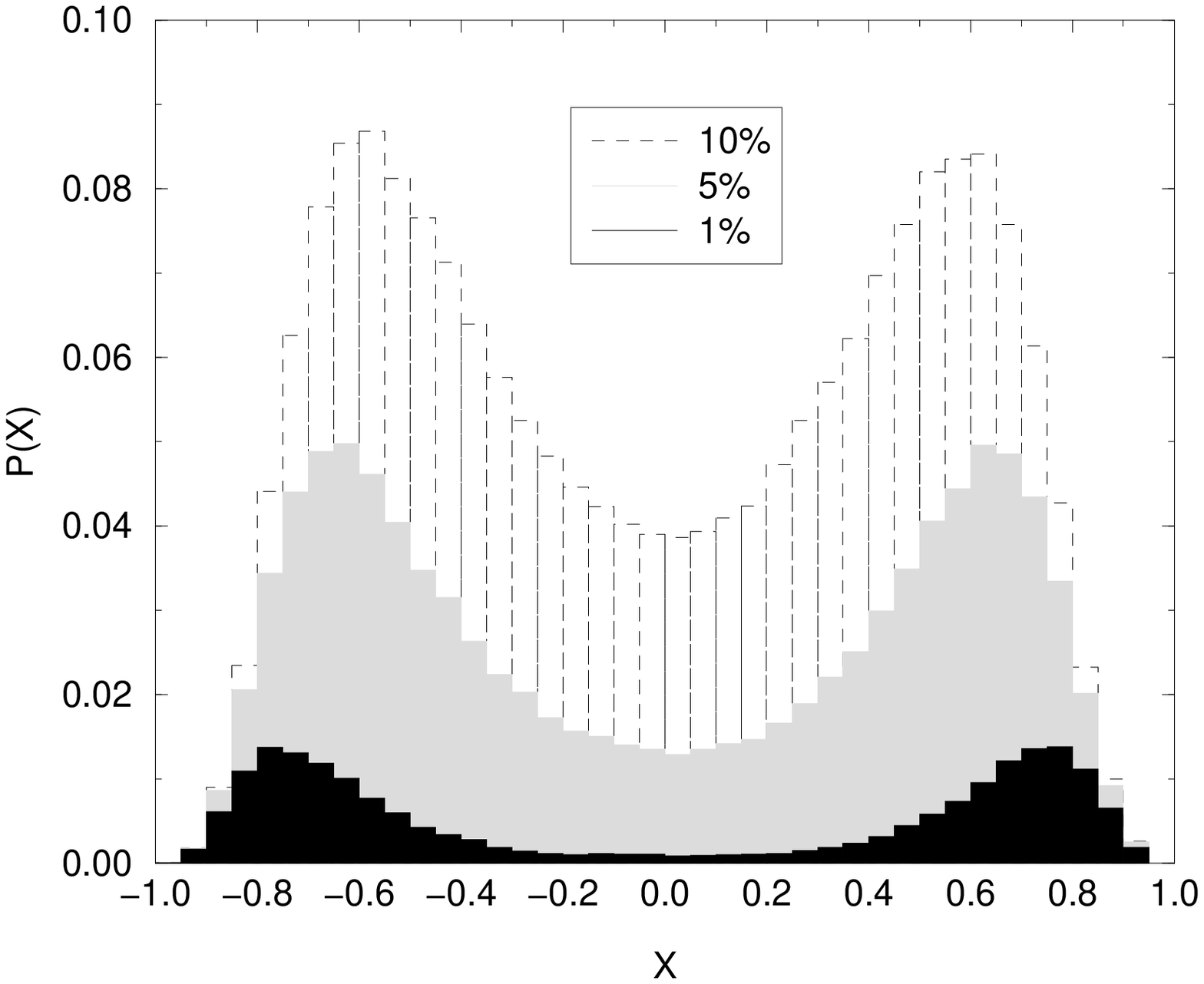} \vskip -6mm
\end{center}
\caption{{}\label{fig:local_chirality} The distribution $P(X)$ 
  for volumes $10^4$ at lattice spacing $0.13\fm$. 
  We include the top 1\%, 5\% and 10\% lattice sites
  with largest density $\Psi^{\dagger} \Psi (x)$. }
\end{figure}

We have analyzed the 10 smallest near-zero modes of the
overlap-improved $D_{\rm ov}^{\rm FP}$ for 60 $10^4$ gauge
configurations at lattice spacing $0.13\fm$. We use a Legendre
polynomial of order 2 to approximate $1/\sqrt{A^{\dagger}A}$ in
$D_{\rm ov}^{\rm FP}$, with the 10 smallest $A^{\dagger}A$ being
projected out and treated exactly. The eigenvalues and eigenvectors
are found using the Arnoldi solver.  In
Fig.~\ref{fig:local_chirality}, we plot the distribution $P(X)$ of the
measure of local chirality $X$ at the lattice sites where the density
$\Psi^{\dagger} \Psi (x)$ of a mode is largest.  The three
distributions correspond to taking, for each mode, 1\%, 5\% and 10\%
of all lattice sites which have the largest density $\Psi^{\dagger}
\Psi (x)$. We do not include exact zero modes, for which $X(x)=\pm 1$
at all lattice sites. We see a very clear double-peaked distribution,
whose maxima are farther from zero if we only include the sites where
the modes are most localized. The maxima are not at $X=\pm 1$ as the
modes are not exactly chiral.  We find the same conclusion as
\cite{DeGrand:2001pj} --- where the near-zero modes are most
localized, they are also very chiral.  Recent work
\cite{Gattringer:2002gn} has shown that at the places where the
near-zero modes of the Dirac operator are concentrated, the gauge
field appears to contain lumps of $F\tilde F$.  This behavior is
consistent with the picture of instanton-dominance of the near-zero
modes, however it is not a conclusive evidence that instantons are the
driving mechanism for chiral symmetry breaking.

\section{Covariant densities and conserved currents}
The GW relation\footnote{The case $2R \neq 1$ in Eq.~\eqref{gw} will
  be considered at the end of this section}
\begin{equation}
\label{D5}
\g D + D \g = D \g D \,.
\end{equation}
implies an exact SU($N_{f}$) $\times$ SU($N_{f}$) global symmetry on
the lattice \cite{Luscher:1998pq}. The vector transformation reads
\begin{equation}
\label{global_V}
\dV^a\psi = i\epsilon T^a \psi \,,
\qquad 
\dV^a\psibar = -i \psibar T^a \epsilon \,,
\end{equation}
while the axial transformation has the form
\begin{equation} \label{global_A}
\dA^a\psi = i\epsilon T^a \gh\psi \,,
\qquad 
\dA^a\psibar = i \psibar \g T^a \epsilon \,,
\end{equation}
where $T^a$, $a=1,\ldots,N_{f}^2-1$ are SU($N_{f}$) generators 
\begin{equation}
[T^a,T^b]=i f_{abc} T^c\,, \qquad {\rm tr}\left( T^a T^b \right) =
\frac{1}{2} \delta^{ab} \,,
\end{equation}
and
\begin{equation} \label{g5hat}
\gh = \g (1-D) \,.
\end{equation}
The action
\begin{equation}
\label{action}
{\cal A} = \overline{\psi} D(U) \psi 
\end{equation}
is invariant under these transformations if $D(U)$ satisfies
Eq.~\eqref{D5}. The fact that ${\cal A}$ is a scalar under the
transformations in Eqs.~\eqref{global_V} and \eqref{global_A} implies
also that it is ${\cal O}(a)$ improved since the mixing of the action
density with other dim=5 operators (in particular, with the clover
term) is forbidden by the symmetries. The spectral quantities are
therefore automatically ${\cal O}(a)$ improved.

The exact global symmetries above imply the existence of conserved
currents. The form of the conserved currents is not unique. It is very
useful to work with conserved currents and scalar and pseudoscalar
densities which transform covariantly (i.e. the same way as in the
formal continuum) under the global transformations in
Eqs.~\eqref{global_V} and \eqref{global_A}.  These dim=3 operators are
again automatically ${\cal O}(a)$ improved: there are no dim=4
operators they can mix with.

Our way to find chiral covariant conserved currents is the same as
that proposed by Kikukawa and Yamada \cite{Kikukawa:1998bg} who
presented explicitly the vector and axial currents in the overlap
construction with Wilson kernel. We present here the currents in the
general case in a form which, we believe, is easy to use in numerical
simulations. We discuss the scalar and pseudoscalar densities, the
Ward identities and the case $2R \neq 1$ also.  In the Appendices we
collect some of the identities implied by the GW relation.

\subsection{A useful form of the currents}
Consider a global transformation $\psi \to \psi+ \delta\psi$, $
\psibar \to \psibar +\delta\psibar $ and assume that the action
Eq.~\eqref{action} is invariant under this transformation, $\delta
{\cal A}=0$. Assume that under the corresponding local transformation
the change of the action can be written in the form
\begin{equation} \label{ADB}
\delta {\cal A} = i \psibar A(U) [ D(U),\epsilon] B(U) \psi \,,
\end{equation}
where, for notational convenience, we treat the infinitesimal,
$x$-dependent parameter of the transformation as a diagonal matrix
\begin{equation}
\left( \epsilon\right)_{xy} = \epsilon(x) \delta_{xy} \,.
\end{equation}
We shall consider the case where $A(U)$ and $B(U)$ transform correctly
under a gauge transformation to make $\delta {\cal A}$ gauge
invariant. The corresponding current is defined through
\begin{equation} \label{dS1}
\delta {\cal A} = i \sum_x \partial_\mu \epsilon(x) J_\mu(x)
= - i \sum_x \epsilon(x)\partial_\mu^* J_\mu(x) \,,
\end{equation}
where $\partial_\mu$ and $\partial_\mu^*$ are the forward and backward
lattice derivatives respectively.

Consider first the case $A=B=1$, the generalization is trivial.  We
extend the gauge fields from $\mathrm{SU}(N_c)$ to
$\mathrm{SU}(N_c)\times\mathrm{U}(1)$ maintaining the gauge covariance
of $D(U)$.  Consider a $\mathrm{U}(1)$ gauge transformation
\begin{equation} \label{U_epsilon}
U_\mu(x)\to\tilde{U}_\mu(x)=\mathrm{e}^{i\epsilon(x)}
U_\mu(x) \mathrm{e}^{-i\epsilon(x+\hat{\mu})} \,.
\end{equation}
For this we have
\begin{equation} \label{Dtilde}
D(U)_{yz} \to D(\tilde{U})_{yz} =
\mathrm{e}^{i\epsilon(y)} D(U)_{yz} \mathrm{e}^{-i\epsilon(z)}\,,
\end{equation}
i.e. for an infinitesimal U(1) gauge transformation the change of
$D(U)$ is given by
\begin{equation}
\dg D(U)= -i [D(U),\epsilon] \,.
\end{equation}
Using Eqs.~\eqref{ADB} and \eqref{dS1} one has
\begin{equation}
\delta {\cal A} = i\sum_x \partial_\mu \epsilon(x) J_\mu(x) =
-\psibar \dg D(U) \psi \,.
\end{equation}
We show below that $\dg D(U)$ can be rewritten in the form
\begin{equation} \label{defK}
\dg D(U)=-i \sum_x \partial_\mu \epsilon(x) K_\mu(x) \,,
\end{equation}
which gives then
\begin{equation}
J_\mu(x)=\psibar K_\mu(x)\psi \equiv
\sum_{yz} \psibar_y (K_\mu(x))_{yz} \psi_z \,. 
\end{equation}
To construct $K_\mu(x)$ satisfying Eq.~\eqref{defK} observe that the
gauge transformation in Eq.~\eqref{Dtilde} can be reached by
performing the changes
\begin{equation} \label{Ualpha}
U_\mu(x) \to U_\mu^{(\alpha)}(x) = 
\mathrm{e}^{i\alpha_\mu(x)}U_\mu(x) 
\end{equation} 
on each link {\em independently} and taking the actual values of
$\alpha_\mu(x)$ to be
\begin{equation}
\alpha_\mu(x) = -\partial_\mu\epsilon(x) \,.
\end{equation}
(Note that the individual changes in Eq.\eqref{Ualpha} are {\em not}
pure gauge transformations -- they only add up to that after all the
links have been properly changed.)

To linear order we have
\begin{equation}
\label{dgD}
\dg D(U)=D(\tilde{U})-D(U) =  -\sum_x \partial_\mu\epsilon(x) 
\left. \frac{\delta D(U_\mu^{(\alpha)})}{\delta \alpha_\mu(x)}
\right|_{\alpha=0} \,.
\end{equation}
Eqs.~\eqref{defK} and ~\eqref{dgD} give the kernel of the current:
\begin{equation} \label{KernelK}
K_\mu(x)=-i\left. \frac{\delta D(U_\mu^{(\alpha)})}{\delta \alpha_\mu(x)}
\right|_{\alpha=0} \,.
\end{equation}

It is easy to see that in the general case of Eq.~\eqref{ADB} the
current is given by the kernel
\begin{equation}
\tilde{K}_\mu(x)= A(U)K_\mu(x;U) B(U) \,,
\end{equation}
where $K_\mu(x)$ is defined in Eq.~\eqref{KernelK}.  Observe that
Eq.~\eqref{KernelK} provides a straightforward way for practical
determination of the kernel $K_\mu(x)$ (and hence of the conserved
currents discussed below) by performing the numerical differentiation
in $\alpha_\mu(x)$.

\subsection{Chiral covariant conserved vector and axial currents}
Consider a global chiral transformation acting only on the
right-handed components $\psi_R=\hat{P}_R \psi$ and
$\psibar_R=\psibar P_L$ :
\begin{equation} \label{global_R}
\dR^a\psi = i\epsilon T^a \hat{P}_R \psi \,,
\qquad 
\dR^a\psibar = -i \psibar P_L T^a \epsilon 
\qquad \text{(global)} \,,
\end{equation}
and the analogous global left-handed transformations
\begin{equation} \label{global_L}
\dL^a\psi = i\epsilon T^a \hat{P}_L \psi \,,
\qquad 
\dL^a\psibar = -i \psibar P_R T^a \epsilon \,,
\qquad \text{(global)} \,.
\end{equation}
The projectors above are defined as
\cite{Niedermayer:1998bi,Narayanan:1998uu}
\begin{equation}
\label{proj1}
\hat{P}_{R/L}=\half (1 \pm \gh)\,,\qquad 
P_{R/L}=\half (1 \pm\g)\,. 
\end{equation}
These transformations are symmetries of the action.  It is convenient
to promote these global transformations to local ones in a way
preserving chirality
\cite{Kikukawa:1998bg},
\begin{equation}
\label{KYR}
\dR^a\psi = 
i T^a \hat{P}_R \epsilon \hat{P}_R  \psi \,,
\qquad 
\dR^a\psibar = 
-i \psibar P_L \epsilon P_L T^a  \,.
\end{equation}
In Eq.~\eqref{KYR} $\epsilon \hat{P}_R \psi$ is not a right-handed
field if $\epsilon$ is $x$-dependent. This explains the presence of
the second projector $\hat{P}_R$ \cite{Kikukawa:1998bg}.
Keeping the form of Eqs.~\eqref{global_R}, \eqref{global_L} in 
the local case would produce conserved currents as well, transforming, 
however, non-covariantly under global transformations. Note that it is
preferable to use the {\em covariant} conserved current. 
For example, the pion is created from the vacuum by the covariant 
current, and the corresponding relation defining $f_\pi$ is
valid for this current. Non-covariant conserved currents result in
an extra $m$-dependent factor (cf. Appendix C).

Using the identity\footnote{This and other useful identities are
  collected in the Appendices.}  $D \hat{P}_R=P_L D$ one can write the
corresponding change of the action as
\begin{equation}
\dR^a {\cal A} =i\psibar \left( 
P_L[D,\epsilon] \hat{P}_R T^a \right) \psi \,.
\end{equation}
This is of the general form considered in Eq.~\eqref{ADB} hence we
readily obtain the corresponding current
\begin{equation}
J_{R\mu}^a(x)=\psibar K_{R\mu}^a(x)\psi\,,
\qquad 
K_{R\mu}^a(x)=P_L K_{\mu}(x) \hat{P}_R T^a \,,
\end{equation}
where the kernel $K_{\mu}(x)$ is defined in Eq.~\eqref{KernelK}.

Similarly, the left-handed local transformation is
\begin{equation}
\label{KYL}
\dL^a\psi = 
i T^a \hat{P}_L \epsilon \hat{P}_L  \psi \,,
\qquad 
\dL^a\psibar = 
-i \psibar P_R \epsilon P_R T^a  \,.
\end{equation}
and the left-handed current is given by
\begin{equation}
J_{L\mu}^a(x)=\psibar K_{L\mu}^a(x)\psi\,,
\qquad 
K_{L\mu}^a(x)=P_R K_{\mu}(x) \hat{P}_L T^a \,.
\end{equation}
Under the global chiral transformations in Eqs.~\eqref{global_R} and
\eqref{global_L} these currents transform covariantly
\begin{equation}
\DR^a J_{R\mu}^b(x) = i f_{abc} J_{R\mu}^c(x) \,,
\qquad
\DL^a J_{L\mu}^b(x) = i f_{abc} J_{L\mu}^c(x) \,.
\end{equation}
The currents $J_{R\mu}$ and $J_{L\mu}$ are, of course, invariant under
the left- and right-handed transformations, respectively.  These
properties imply that the vector and axial currents
\begin{equation} \label{Vmua}
\begin{split}
V_\mu^a(x)=J_{R\mu}^a(x)+J_{L\mu}^a(x) 
& = \psibar \left( P_L K_{\mu}(x) \hat{P}_R + P_R K_{\mu}(x) \hat{P}_L
\right) T^a\psi \\
& = \half \psibar \left( K_{\mu}(x) - \g K_{\mu}(x) \gh \right) T^a\psi \,,
\end{split}
\end{equation}
\begin{equation} \label{Amua}
\begin{split}
A_\mu^a(x)=J_{R\mu}^a(x)-J_{L\mu}^a(x)
& = \psibar \left( P_L K_{\mu}(x) \hat{P}_R - P_R K_{\mu}(x) \hat{P}_L
\right) T^a\psi \\
& = \half \psibar \left( - \g K_{\mu}(x) + K_{\mu}(x) \gh \right) T^a\psi \,,
 \end{split}
\end{equation}
are transformed under global vector and axial rotations
\begin{equation}
\label{KYVA}
\dV^a=\dR^a + \dL^a\,, \qquad \dA^a=\dR^a - \dL^a\,,
\end{equation}
covariantly
\begin{equation}
\begin{split}
  & \DV^a V_\mu^b(x)= i f_{abc} V_\mu^c(x)\,, \quad
  \DV^a A_\mu^b(x)= i f_{abc} A_\mu^c(x)\,, \quad \\
  & \DA^a V_\mu^b(x)= i f_{abc} A_\mu^c(x)\,, \quad \DA^a A_\mu^b(x)=
  i f_{abc} V_\mu^c(x)\,.
\end{split}
\end{equation}

Let us discuss the overhead related to the currents in
Eqs.~\eqref{Vmua} and \eqref{Amua} in comparison with using the
non-conserved, non-covariant local currents. The action of the kernel
$K_\mu(x)$ on a vector $v$ is $\propto D(U^{(\alpha)})v-D(U)v$ in the
simplest numerical approximation of the derivative in
Eq.~\eqref{KernelK}. The projectors $\hat{P}_R,\hat{P}_L$ require an
additional multiplication with $D(U)$. We believe, this small overhead
is justified by the advantages of having conserved and ${\cal O}(a)$
improved currents.

Note that the fermion fields transform under the local vector 
and axial transformations as
\begin{equation}
\label{VTR}
\dV^a\psi = 
i T^a \half \left( \epsilon + \gh \epsilon \gh \right)  \psi \,,
\qquad 
\dV^a\psibar = -i \psibar \epsilon T^a \,,
\end{equation}
\begin{equation}
\label{ATR}
\dA^a\psi = 
i T^a \half \left( \gh \epsilon + \epsilon \gh \right)  \psi \,,
\qquad 
\dA^a\psibar = i \psibar \epsilon \g T^a \,.
\end{equation}

\subsection{Chiral covariant scalar and pseudoscalar densities}
It is easy to show that the scalar and pseudoscalar quantities
\begin{equation}
\label{SP}
S^a=\psibar \left( 1-\frac{1}{2}D \right) T^a \psi \,,
\qquad
P^a=\psibar \g \left( 1-\frac{1}{2}D \right) T^a \psi \,,
\end{equation}
$a=0,1,\dots N_f^2-1$ (with $T^0=1$) transform under global vector
and axial rotations like in the formal continuum,
\begin{equation}
\DV^a S^b = i f_{abc} S^c\,, \qquad
  \DV^a P^b = i f_{abc} P^c \,,
\end{equation}
\begin{equation}
\DA^a S^b = - d_{abc} P^c\,, \qquad
  \DA^a P^b = - d_{abc} S^c \,,
\end{equation}
where $\{T^a,T^b\}=d_{abc} T^c$. In particular, for a flavour singlet
axial transformation we have
\begin{equation}
 \DA S^a =-2 P^a \,, \qquad
 \DA P^a =-2 S^a \,,
\end{equation}
while the non-singlet axial transformation of the flavour singlet
quantities reads
\begin{equation}
 \DA^a S = -2 P^a \,, \qquad
 \DA^a P = -2 S^a \,, \qquad a=1,\dots N_f^2-1\,.
\end{equation}
(Here $\dA \equiv \dA^0$, $S=S^0$, etc. for the flavour singlet
quantities.)

Since the quantity $S$ enters the action in the mass term $m S$, 
we need the variation of $S$ under a {\em local} chiral transformation 
when considering  Ward identities.
They read
\begin{equation}
i\frac{\delta_A^a S}{\delta\epsilon(x)} = -2 P^a(x)\,, \qquad
i\frac{\delta_A^a P}{\delta\epsilon(x)} = -2 S^a(x)\,,
\end{equation}
where $S^a(x)$ and $P^a(x)$ are the covariant scalar and pseudoscalar 
densities related to the divergencies of conserved covariant currents
\eqref{Vmua}, \eqref{Amua}.
Using Eqs.~\eqref{KYR},\eqref{KYL},\eqref{KYVA} one obtains
\begin{equation} \label{Sax} 
\begin{split}
S^a(x) = \psibar 
  & \left[ \half E(x) \left(1-\half D\right) \right. \\
  & + \left. \frac{1}{4} \left(1-\half D\right)  
     \left( E(x) + \gh E(x) \gh \right) \right] T^a \psi \,,
\end{split}
\end{equation}
\begin{equation} \label{Pax}
\begin{split}
P^a(x) = \psibar 
  & \left[ \half E(x) \g \left(1-\half D\right) \right.  \\
  & + \left. \frac{1}{4} \left(1-\half D\right) 
      \left( \gh E(x) + E(x) \gh\right) \right] T^a \psi \,.
\end{split}
\end{equation}
Here we have introduced the notation
\begin{equation} \label{Ex} 
\left( E(x) \right)_{yz} = \delta_{xy} \delta_{xz} \,.
\end{equation}

When summed over the lattice these densities reproduce the quantities
in Eq.~\eqref{SP}:
\begin{equation}
\sum_x S^a(x) =S^a \,,\qquad
\sum_x P^a(x) =P^a \,,
\end{equation}
as it should be since $\sum_x \delta/\delta\epsilon(x)$ corresponds to
an infinitesimal {\em global} transformation.

Although Eqs.~\eqref{Sax},\eqref{Pax} look inconveniently complicated, 
up to contact terms and prefactors $(1+ {\rm O}(am))$ 
they can be replaced by the corresponding point-like
densities (cf. Appendix C).

Note that under a global transformation the first and second terms 
in Eqs.~\eqref{Sax}, \eqref{Pax} transform separately, 
i.e. the simpler expressions
\begin{equation} \label{SP_simple} 
 \psibar E(x) \left(1-\half D\right) T^a \psi \,,
\qquad
 \psibar E(x) \g \left(1-\half D\right) T^a \psi
\end{equation}
also define covariant densities.
Similarly, the non-conserved currents
\begin{equation} \label{VA_simple} 
 \psibar E(x) \gamma_\mu \left(1-\half D\right) T^a \psi \,,
\qquad
 \psibar E(x) \gamma_\mu \g \left(1-\half D\right) T^a \psi
\end{equation}
are also transforming covariantly. 
They are not related, however, to each other by a Ward identity.

\subsection{Ward identities}
Consider the fermion action with a flavour invariant mass
term\footnote{{}In our convention the Boltzmann factor 
is $\exp(-{\cal A}_m)$.}
\begin{equation}
\begin{split}
  {\cal A}_m & = \psibar D\psi + m(\psibar_L \psi_R + \psibar_R \psi_L) \\
  & = {\cal A} + m S = \psibar \left( D + m\left(1-\frac{1}{2}D\right)
  \right) \psi \,.
\end{split}
\end{equation}
Under a local axial transformation defined in
Eqs.~\eqref{KYR},\eqref{KYL} and \eqref{KYVA} we get
\begin{equation}
i\frac{\dA^a{\cal A}_m}{\delta\epsilon(x)} = 
\partial_\mu^{*} A_\mu^a(x) -2m P^a(x) \,.
\end{equation} 
Consider the local Ward identity obtained by the change of variables
defined by an axial transformation in the path integral for the
fermionic expectation value of a multi-local operator
$\cO(y_1,y_2,\dots,y_n)$:
\begin{equation} \label{localWI}
\left\langle i\frac{\dA^a\cO}{\delta\epsilon(x)}\right\rangle_{\rm F}
- \left\langle \cO \partial_\mu^* A_\mu^a(x) \right\rangle_{\rm F}
+ 2m \left\langle \cO P^a(x) \right\rangle_{\rm F}
-\delta^{a0} 2N_f \left\langle \cO q(x) \right\rangle_{\rm F} =0 \,,
\end{equation}
where $P(x)$ is given by Eq.~\eqref{Pax} and the
un-normalized fermionic expectation value (in a given background gauge
field) is defined as
\begin{equation}
\langle {\cal O} \rangle_{\rm F} \equiv
\int [d\psibar d\psi] {\rm e}^{-\psibar D \psi} {\cal O}(\psibar,\psi) \,.
\end{equation}
The last term in Eq.~\eqref{localWI} is the contribution from the
measure which is not invariant under a flavour singlet axial
transformation \cite{Luscher:1998pq}, and its change is given by
\begin{equation}
[d\psi'] \to [d\psi']=[d\psi](1+\delta\mu)
\end{equation}
where\footnote{{}To get the correct sign keep in mind that $\psibar$,
  $\psi$ are Grassmann variables.}
\begin{equation}
\label{measure} 
\delta\mu =-i {\rm Tr}\left( \hPR\epsilon\hPR - \hPL\epsilon\hPL\right)
 = -i {\rm Tr}\left(\gh  \epsilon\right)
 = 2 i N_f \sum_x \epsilon(x) q(x) \,,
\end{equation}
or
\begin{equation}
i\frac{\dA^a\mu}{\delta\epsilon(x)}=-2N_f\, q(x) \,.
\end{equation}
(Note that the $[d\psibar]$ part of the measure is invariant.) In
Eq.~\eqref{measure} $q(x)$ is the topological charge density
\begin{equation} \label{qtop}
q(x) = \frac{1}{2}{\rm tr}(\g D(x,x)) \,,
\end{equation}
which enters the index theorem \cite{Hasenfratz:1998ri}
\begin{equation}
\sum_x {\rm tr}(\g D(x,x)) = N_f ~{\rm index}(D) \,.
\end{equation}
If $x$ in Eq.~\eqref{localWI} is sufficiently far removed from the
operator $\cO(y_1,y_2,\dots,y_n)$ ($|y_i-x|$ is much larger than the
range of $D$), the first term in Eq.~\eqref{localWI} is zero. In this
case the Ward identity is consistent with the classical equations of
motion
\begin{equation}
\partial_\mu^* A_\mu^a(x) = 2m P^a(x)
\end{equation} 
for $a \ne 0$. Summing over $x$ in Eq.~\eqref{localWI} leads to the
global Ward identity
\begin{equation}
\label{globalWI}
\left\langle i \frac{\dA^a {\cal O}}{\delta\epsilon} \right\rangle_{\rm F}
 + 2m\left\langle  {\cal O} P^a \right\rangle_{\rm F} 
 - \delta^{a0} 2\nu N_f \left\langle {\cal O} \right\rangle_{\rm F} =0 \,,
\end{equation}
where $\nu$ is the value of the topological charge of the given gauge
configuration.

Let us illustrate the consequences of Eq.~\eqref{globalWI} on two
examples. Consider first $a=0$, $\cO=1$:
\begin{equation} \label{mP}
m\langle P \rangle_{\rm F} = 
\nu N_f \langle 1 \rangle_{\rm F} \,.
\end{equation}
Combining this relation with that obtained by setting $\cO=P$:
\begin{equation}
m \langle P P \rangle_{\rm F} - \langle S \rangle_{\rm F}
-\nu N_f \langle P \rangle_{\rm F} =0 \,.
\end{equation}
we get
\begin{equation}
\nu^2 N_f^2 \langle 1 \rangle_{\rm F} = 
m^2 \langle P P \rangle_{\rm F} - m \langle S \rangle_{\rm F} \,.
\end{equation}
Averaging over the gauge fields leads to an identity for the
topological susceptibility
\begin{equation}
\chi_{\rm top} \equiv \frac{ \langle \nu^2\rangle}{V}
= -\frac{m}{N_f^2} \langle S(0) \rangle +
\frac{m^2}{N_f^2} \frac{1}{V} \langle P P \rangle \,.
\end{equation}
Assuming that there exists no massless excitation in the flavour
singlet chanel we get in the chiral limit
\cite{Chandrasekharan:1999wg}
\begin{equation}
\frac{ \langle \nu^2\rangle}{V}=\frac{m}{N_f}\Sigma \,, \qquad m
\rightarrow 0 \,.
\end{equation}
In the second example consider the non-singlet Ward identity with $\cO
= P^b$
\begin{equation}
2 m\left\langle P^a P^b \right\rangle =
 d_{abc}\left\langle S^c \right\rangle \,,
\end{equation}
and in particular
\begin{equation}
2m\left\langle P^1 P^1 \right\rangle =
 \frac{1}{N_f}\langle S \rangle \,.
\end{equation}
(Note that $d_{110}=1/N_f$). In the chiral limit the pseudoscalar
correlator is saturated by the Goldstone boson pole, while
\begin{equation}
\langle S \rangle = V \langle S(0) \rangle = - V \Sigma N_f
\end{equation}
leading to the GMOR relation on the lattice \cite{Hasenfratz:1998bb}
\begin{equation} \label{GMOR}
f_\pi^2 m_\pi^2 = 4m\Sigma \,.
\end{equation}

\subsection{{}The case of the general GW relation ${\mathbf 2R \ne 1}$ }
Consider the Dirac operator satisfying the general GW relation
\begin{equation} \label{GWR}
\g D + D\g =  D \g 2R D \,.
\end{equation}

In order to connect this with the $2R \to 1$ case we rescale D and the
field variables:
\begin{equation} \label{DctoD}
D=\left( 2R\right)^{-1/2} D_1
\left(2R\right)^{-1/2} \,,
\end{equation}
and
\begin{equation}
\label{redef}
\psibar=\psibar_1 \left( 2R\right)^{1/2} \,,
\quad
\psi= \left( 2R\right)^{1/2} \psi_1 \,.
\end{equation}
Obviously, we have
\begin{equation} \label{Act_mod}
{\cal A}(m)  =
\psibar_1 \left[ D_1 + m \left( 1 - \half D_1 \right) \right] \psi_1 
 =\psibar D(m) \psi \,, 
\end{equation}
where
\begin{equation} \label{Dm_mod0}
D(m)  = D +m \left( \RRi - \half D \right) \,.
\end{equation}

The new Dirac operator $D_1$ satisfies the GW relation with $2R=1$
\begin{equation} \label{GWc}
\g D_1 + D_1 \g =  D_1 \g D_1 \,.
\end{equation}
The spectrum of $D_1$ lies on a circle of radius $1$ with the centre
at $1$.  
Assuming that $D$ is normalized in the standard way,
so that for the free case its FT is 
$\tilde{D}(p)= i\gamma_\mu p_\mu +{\cal O}(ap^2)$,
the operator $D_1$ will be normalized differently:
$\tilde{D}_1(p)=c i\gamma_\mu p_\mu + {\cal O}(ap^2)$ with
$c=2\tilde{R}(0)$.  This could be restored by a simple rescaling of
the fields, however, for convenience we shall rather keep the
non-conventional normalization of $D_1$.  Of course, this choice does
not affect the physical results.

One can derive currents and densities by writing them first in terms
of $\psi_1$, $\psibar_1$ and $D_1$. 
We choose here, however, a more direct way, and work with the
original variables $\psi$, $\psibar$ and $D$.
To start with, we introduce the operators
\begin{align}
\label{proj}
\Gh &=
(2R)^{1/2}\gh(2R)^{-1/2}=\g(1-2RD)\,, \\
\hat{{\cal P}}_R &= \frac{1}{2}(1 + \Gh)\,, \qquad
\hat{{\cal P}}_L=\frac{1}{2}(1 - \Gh)\,, \notag
\end{align}
which are related to their counterparts at $2R=1$ by a 
{\it similarity} transformation. Although, these operators are 
not hermitian, their basic properties are unchanged:
\begin{equation}
\g D = - D \Gh \,, \quad
\Gh^2=1\,, \quad
\hat{{\cal P}}_R \hat{{\cal P}}_R = \hat{{\cal P}}_R \,, \quad
\hat{{\cal P}}_R \hat{{\cal P}}_L =0\,, \quad
\hat{{\cal P}}_R + \hat{{\cal P}}_L =1\,.
\end{equation}

It is convenient to define the local transformations through
(cf. \eqref{VTR}, \eqref{ATR})
\begin{equation}
\label{VTR_mod}
\dV^a\psi = 
i T^a \half \left( \epsilon + \Gh \epsilon \Gh \right)  \psi \,,
\qquad 
\dV^a\psibar = -i \psibar \epsilon T^a \,,
\end{equation}
\begin{equation}
\label{ATR_mod}
\dA^a\psi = 
i T^a \half \left( \Gh \epsilon + \epsilon \Gh \right)  \psi \,,
\qquad 
\dA^a\psibar = i \psibar \epsilon \g T^a \,.
\end{equation}
The corresponding global transformation is a symmetry for the massless
case.
The conserved currents are given by expressions analogous to 
Eqs.~\eqref{Vmua}, \eqref{Amua} with the replacements
$\hat{P}_{R/L} \to \hat{\cal{P}}_{R/L}$, $\gh \to \Gh$, and where
$K_\mu(x)$ is also given by Eq.~\eqref{KernelK}.
It is straightforward to verify that they are transforming covariantly
under global transformations.

Similarly, the covariant densities obtained from 
\begin{equation}
S = \psibar \left( \RRi - \half D \right) \psi
\end{equation}
are given by 
\begin{equation} \label{Sax_mod} 
\begin{split}
S^a(x) = \psibar 
  & \left[ \half E(x) \left(\RRi - \half D\right) \right. \\
  & + \left. \frac{1}{4} \left(\RRi - \half D\right)  
     \left( E(x) + \Gh E(x) \Gh \right) \right] T^a \psi \,,
\end{split}
\end{equation}
\begin{equation} \label{Pax_mod}
\begin{split}
P^a(x) = \psibar 
  & \left[ \half E(x) \g \left(\RRi - \half D\right) \right.  \\
  & + \left. \frac{1}{4} \left(\RRi - \half D\right) 
      \left( \Gh E(x) + E(x) \Gh\right) \right] T^a \psi \,.
\end{split}
\end{equation}

The presence of $2R \ne 1$ leads to a small overhead only: $2R$ and
$(2R)^{-1}$ are local, non-singular operators without Dirac indices.
Since the inverse of $D(m)$ can be written as 
\begin{equation}
\label{Dm_inverse}
D(m)^{-1}=\frac{1}{1-m/2} 2R \left[ D 2R + \frac{m}{1-m/2}\right]^{-1} \,,
\end{equation}
the multi-mass solver can be easily generalized to this case. The
eigenvalue equation for $D_1$ leads to the generalized eigenvalue
equation for $D$
\begin{equation}
D |\phi_\lambda \rangle= \lambda \RRi |\phi_\lambda \rangle \,,
\end{equation}
where the eigenvector $|\phi_\lambda \rangle$ is related to that of
$D_1$ by
\begin{equation}
|\phi_\lambda \rangle= (2R)^{1/2} |\phi^1_\lambda \rangle\,,
\end{equation}
and forms an ortho-normalized system with the weight $(2R)^{-1}$ 
\begin{equation}
\langle\phi_\lambda ' |\RRi |\phi_\lambda \rangle=
\delta_{\lambda \lambda '}\,.
\end{equation}

Note finally that the presence of the extra factors appearing in 
eqs.~\eqref{Sax_mod},\eqref{Pax_mod} in fact simplifies the
calculation of expectation values because due to the identity
(cf. \eqref{Dm_mod})
\begin{equation}
\left( \RRi -\frac{1}{2}D \right) D(m)^{-1} = 
\frac{1}{(1-m/2)^2} \left[ D 2R + \frac{m}{1-m/2}\right]^{-1}
 - \frac{1}{2-m} 
\end{equation}
it removes the factor $2R$ in \eqref{Dm_inverse}.

One can show that the covariant densities appearing in 
the Ward identities can be replaced also in this case 
(up to contact terms and $am$ dependent factors going to 1 in 
the continuum limit) by simpler operators (cf. Appendix C)
\begin{equation}
\half \psibar \left( E(x) \RRi + \RRi E(x) \right)\psi \,, \qquad 
\half \psibar \left( E(x) \RRi + \RRi E(x) \right) \g \psi \,.
\end{equation}

\section{Conclusions}
The main purpose of this study has been to show that it is feasible to
use the parametrized fixed-point QCD action in simulations.  We also
give a practical and general construction of conserved currents and
covariant densities for chiral lattice actions.  The tests of the
parametrized fixed-point Dirac operator $D^{\rm FP}$ show that the
deviations from chiral symmetry are small and can be removed by small
corrections in a straightforward fashion with the overlap
construction. A first study of the hadron spectroscopy shows that the
additive quark mass renormalization and, more importantly, the
fluctuation of it, are small, allowing us to go quite small physical
quark masses. The speed of light, extracted from the
momentum-dependence of the light hadron spectrum, is consistent with
1, evidence that the fixed-point properties are intact.  We also
measure the renormalized chiral condensate directly using
finite-volume scaling, giving 
$\Sigma_{\overline{\rm MS}}(2\GeV)=(262 \pm 12\MeV)^3$, 
which is in good agreement with
other recent measurements. In addition, we measure the quenched
topological susceptibility $\chi_t=(196 \pm 6\MeV)^4$ and test
other methods of determining $\Sigma$, using the pion mass
measurements or measuring the distribution of the smallest non-zero
eigenvalue of the Dirac operator. We also examine near-zero modes of
the Dirac operator and find that they do appear to be chiral locally,
as other studies have found, in support of the picture of
instanton-dominance.

Any chiral lattice action is much more expensive to use in simulations
than the standard actions. How competitive is the fixed-point QCD
action with, say, the overlap operator? The majority of the simulation
time is spent inverting the Dirac operator, so using the more
expensive fixed-point gauge action $S^{\rm FP}_g$, with its desirable
properties, is a small part of the overall cost. If one is interested
in observables where small deviations from chiral symmetry are
acceptable, the parametrized operator $D^{\rm FP}$ can be used.  An
optimal implementation of $D_{\rm ov}$ with $D^{\rm W}$ in the kernel,
say the rational approximation method, which is on the order of
100-200 times more costly than using $D^{\rm W}$, has chiral symmetry
violations which are orders of magnitude smaller. The overlap-improved
$D^{\rm FP}_{\rm ov}$, constructed with a low order polynomial,
achieves the same accuracy of chiral symmetry at a similar cost.
Comparison of overlap and domain wall fermions does not show a large
difference in cost for a given accuracy of chiral symmetry
\cite{Hernandez:2000iw}.  The major advantage of the fixed-point
action is if the lattice spacing dependence is much reduced. If this
is so, the cost of using the fixed-point action is offset by being
able to work on much coarser lattices. A large scale and systematic
study, which is in progress, is required to accurately determine how
large the lattice artifacts are.

Quenched QCD simulations with chiral fermions have advanced very
quickly over the last few years. However, the serious problem of how
to implement chiral fermions in full QCD will have to be tackled.  It
is also an open question, how necessary chiral lattice fermions are
for much of QCD phenomenology.

{\bf Acknowledgements:} The authors are indebted to Gilberto Colangelo, 
Tom DeGrand, J\"urg Gasser, Christof Gattringer, Leonardo Giusti,
Maarten Golterman, Jimmy Juge, Julius Kuti, Christian Lang, Steve Sharpe
and Peter Weisz for useful discussions.  
We also thank the members of the BGR collaboration for their support.

This work has been supported by the Schweizerischer Nationalfonds, by
the US DOE under grant DOE-FG03-97ER40546 and by the European
Community's Human Potential Programme under contract
HPRN-CT-2000-00145.

\newpage

\setcounter{section}{0}
\newappendix{}
Some of the useful identities, which follow from the Ginsparg-Wilson
relation are collected here (case $2R = 1$). Concerning the
definitions of $\PR,\PL,\gh,\hPR$ and $\hPL$ we refer to
Eqs.~\eqref{g5hat}, \eqref{proj1}.

\begin{equation}
\g D = - D\gh   \,,
\end{equation}


\begin{equation}
D\hPR=\PL D\,, \qquad D\hPL=\PR D\,,
\end{equation}

\begin{equation}
D = \PR D \hPL +  \PL D \hPR \,,
\end{equation}

\begin{equation}
\begin{split}
  & 2\PR\hPL=\PR D \hPL = \PR D = D \hPL \,, \\
  & 2\PL\hPR=\PL D \hPR = \PL D = D \hPR \,,
\end{split}
\end{equation}

\begin{equation}
\PR\hPR = 
\PR \left( 1-\frac{1}{2}D\right) = \left( 1-\frac{1}{2}D\right) \hPR \,,
\end{equation}

\begin{equation}
\PL\hPL = 
\PL \left( 1-\frac{1}{2}D\right) = \left( 1-\frac{1}{2}D\right) \hPL \,,
\end{equation}

\begin{equation}
\PR\hPR + \PL\hPL = 1-\frac{1}{2} D \,,
\end{equation}

\begin{equation}
\PR\hPR - \PL\hPL = 
\g \left( 1-\frac{1}{2}D\right) = \left( 1-\frac{1}{2}D\right) \gh \,,
\end{equation}

\begin{equation} \label{Dm_rel}
D(m) = D  + m \left( 1-\half D \right)\,, \qquad
1-\half D = \frac{1}{1-m/2}\left( 1-\half D(m)\right) \,.
\end{equation}

\newappendix{}
We consider here identities which follow from the general GW relation
with $2R \ne 1$ . Concerning the definitions of $\Gh,\hcPR$ and
$\hcPL$ we refer to Eq.~\eqref{proj}.

\begin{equation}
\g D = - D\Gh \,,
\end{equation}

\begin{equation}
D\hcPR=\PL D\,, \qquad D\hcPL=\PR D\,,
\end{equation}

\begin{equation}
D = \PR D \hcPL +  \PL D \hcPR \,,
\end{equation}

\begin{equation}
\begin{split}
  & \RRi 2\PR\hcPL=\PR D \hcPL = \PR D = D \hcPL \,, \\
  & \RRi 2\PL\hcPR=\PL D \hcPR = \PL D = D \hcPR \,,
\end{split}
\end{equation}

\begin{equation}
\g \left( \RRi -\half D\right) = \left( \RRi - \half D\right) \Gh \,,
\end{equation}

\begin{equation}
\RRi \PR \hcPR = 
\PR \left( \RRi - \half D\right) = \left( \RRi -\half D\right) \hcPR \,,
\end{equation}

\begin{equation}
\RRi \PL \hcPL =
\PL \left( \RRi - \half D\right) = \left( \RRi - \half D\right) \hcPL \,,
\end{equation}

\begin{equation}
\RRi \left( \PR\hcPR + \PL\hcPL\right) = \RRi - \half D \,,
\end{equation}

\begin{equation} \label{Dm_mod}
D(m) = D  + m \left( \RRi - \half D \right)\,, \quad
\RRi - \half D = \frac{1}{1-m/2}\left( \RRi - \half D(m)\right) .
\end{equation}

\newpage

\newappendix{}
We show here, first for the case $2R=1$, that up to contact 
terms\footnote{{}To be understood in a broader sense: 
  not necessarily proportional to $\delta_{xy}$ but negligible when 
  $|x-y|$ is larger than the range of the Dirac operator.} 
the covariant densities \eqref{Sax}, \eqref{Pax} can be replaced
by their point-like counterparts.

According to Eqs.~\eqref{Pax}, \eqref{Dm_rel} one has
\begin{equation} \label{Pax1}
\begin{split}
P^a(x) = \frac{1}{1-m/2}
\psibar & \left[ \half E(x) \g \left(1-\half D(m)\right) \right.   \\
+ & \left. \frac{1}{4} \left(1-\half D(m)\right) 
\left( \gh E(x)+  E(x) \gh\right) \right] T^a \psi \,.
\end{split}
\end{equation}
In the matrix elements one has to calculate $D(m)^{-1}[ \ldots ] D(m)^{-1}$
and the terms proportional to $D(m)$ can be omitted:
\begin{equation} \label{Pax2}
P^a(x) \approx \frac{1}{1-m/2}\psibar \left[ 
\half E(x) \g +\frac{1}{4}\gh E(x) + \frac{1}{4} E(x) \gh \right] T^a \psi \,,
\end{equation}
where the $\approx$ sign means ``equivalent up to contact terms''.
The following relations can be easily verified
\begin{equation} \label{gh_rel}
\gh = \frac{1-m/2}{1+m/2} \g + \frac{1}{1+m/2} D(m) \gh =
  \frac{1+m/2}{1-m/2} \g - \frac{1}{1-m/2} \g D(m) \,.
\end{equation}
For the $\gh E(x)$ term in \eqref{Pax2} we use the first relation
while for $E(x) \gh$ the second one. 
In this case one of the propagators is eliminated by the corresponding 
$D(m)$, leading again to contact terms. The final result is
\begin{equation} \label{P_to_p1}
\begin{split}
P^a(x) & \approx \frac{1}{(1-m/2)(1-m^2/4)} \psibar E(x) \g T^a \psi \\
 & \approx \frac{1}{1-m^2/4}\psibar E(x)\g 
   \left( 1-\half D\right) T^a\psi\,.
\end{split}
\end{equation}
Note that these prefactors are $1+{\rm O}(am)$ (restoring the lattice 
spacing $a$), i.e. even omitting them does not affect the continuum limit.

The analogous calculation for the scalar density gives
\begin{equation} \label{S_to_s1}
S^a(x) \approx \frac{1}{1-m/2} \psibar E(x) T^a \psi
\approx \psibar E(x)\left( 1-\half D\right) T^a\psi\,.
\end{equation}

Measuring the correlator of the covariant pseudoscalar density 
one can directly determine the pion decay constant $f_\pi$
from the relation
\begin{equation} \label{f_pi}
\sum_{\vec{x}}\langle 0 | P^a(\vec{x},t) P^b(0,0)| 0 \rangle 
  = - \frac{f_\pi^2 m_\pi^3}{16 m^2} \delta^{ab} {\rm e}^{-m_\pi |t|}  \,.
\end{equation}
(The normalization corresponds to the GMOR relation \eqref{GMOR},
$f_\pi^2 m_\pi^2 = 4 m \Sigma$, i.e. $f_\pi \approx 130 \MeV$.)

Finally, we give the generalization of Eqs.~\eqref{P_to_p1},
\eqref{S_to_s1} for the case $2R \ne 1$:
\begin{equation} \label{P_to_p1a}
P^a(x) \approx \frac{1}{1-m/2} \cdot \half
\psibar \left[ E(x) \RRi + \frac{1+m^2/4}{1-m^2/4} \RRi E(x)\right] 
   \g T^a \psi \,,
\end{equation}
and
\begin{equation} \label{S_to_s1a}
S^a(x) \approx \frac{1}{1-m/2} \cdot \half
\psibar \left[ E(x) \RRi + \RRi E(x)\right] T^a \psi \,.
\end{equation}

\eject


\newpage

\newpage


\begin{thebibliography}{99}
  
\bibitem{Kaneko:2001ux}
  Some recent summaries: \\
  T.~Kaneko,
  Nucl.\ Phys.\ Proc.\ Suppl.\  {\bf 106}, 133 (2002); \\
  V.~Lubicz,
  Nucl.\ Phys.\ Proc.\ Suppl.\  {\bf 94}, 116 (2001); \\
  S.~M.~Ryan,
  Nucl.\ Phys.\ Proc.\ Suppl.\  {\bf 106}, 86 (2002); \\
  F.~Karsch,
  Nucl.\ Phys.\ Proc.\ Suppl.\  {\bf 83}, 14 (2000); \\
  L.~Lellouch, Nucl.\ Phys.\ Proc.\ Suppl.\ {\bf 94}, 142 (2001).
  
\bibitem{Hasenfratz:2000sa} P.~Hasenfratz and F.~Niedermayer, Nucl.\ 
  Phys.\ B {\bf 596}, 481 (2001);
  {\tt arXiv:hep-lat/0112003}; \\
  M.~Hasenbusch, P.~Hasenfratz, F.~Niedermayer, B.~Seefeld and
  U.~Wolff, Nucl.\ Phys.\ Proc.\ Suppl.\ {\bf 106}, 911 (2002).
  
\bibitem{Nielsen:1980rz} H.~B.~Nielsen and M.~Ninomiya, Nucl.\ Phys.\ 
  B {\bf 185}, 20 (1981).
  
\bibitem{Kaplan:1992bt} D.~B.~Kaplan,
  Phys.\ Lett.\ B {\bf 288}, 342 (1992); \\
  Y.~Shamir,
  Nucl.\ Phys.\ B {\bf 406}, 90 (1993); \\
  V.~Furman and Y.~Shamir, Nucl.\ Phys.\ B {\bf 439}, 54 (1995).
  
\bibitem{Narayanan:ss} R.~Narayanan and H.~Neuberger, Phys.\ Rev.\ 
  Lett.\ {\bf 71}, 3251 (1993);
  Nucl.\ Phys.\ B {\bf 412}, 574 (1994);
  Nucl.\ Phys.\ B {\bf 443}, 305 (1995); \\
  S.~Randjbar-Daemi and J.~Strathdee, Phys.\ Lett.\ B {\bf 348}, 543
  (1995);
  Nucl.\ Phys.\ B {\bf 443}, 386 (1995);
  Nucl.\ Phys.\ B {\bf 466}, 335 (1996);
  Phys.\ Lett.\ B {\bf 402}, 134 (1997).
  
\bibitem{Hasenfratz:1993sp} P.~Hasenfratz and F.~Niedermayer,
  Nucl.\ Phys.\ B {\bf 414}, 785 (1994); \\
  U.~J.~Wiese, Phys.\ Lett.\ B {\bf 315}, 417 (1993);\\
  T.~DeGrand, A.~Hasenfratz, P.~Hasenfratz and F.~Niedermayer, Nucl.\ 
  Phys.\ B {\bf 454}, 587 (1995);
  Nucl.\ Phys.\ B {\bf 454}, 615 (1995);
  Phys.\ Lett.\ B {\bf 365}, 233 (1996); \\
  W.~Bietenholz, R.~Brower, S.~Chandrasekharan and U.~J.~Wiese,
  Nucl.\ Phys.\ Proc.\ Suppl.\  {\bf 53}, 921 (1997); \\
  T.~DeGrand, A.~Hasenfratz, P.~Hasenfratz, P.~Kunszt and
  F.~Niedermayer,
  Nucl.\ Phys.\ Proc.\ Suppl.\  {\bf 53}, 942 (1997); \\
  K.~Orginos, W.~Bietenholz, R.~Brower, S.~Chandrasekharan and
  U.~J.~Wiese, Nucl.\ Phys.\ Proc.\ Suppl.\ {\bf 63}, 904 (1998).
  
\bibitem{Ginsparg:1981bj} P.~H.~Ginsparg and K.~G.~Wilson, Phys.\ 
  Rev.\ D {\bf 25}, 2649 (1982).
  
\bibitem{Hasenfratz:1997aa} P.~Hasenfratz,
  Nucl.\ Phys.\ Proc.\ Suppl.\  {\bf 63}, 53 (1998);\\
  H.~Neuberger, Phys.\ Lett.\ B {\bf 427}, 353 (1998);\\
  Y.~Kikukawa and T.~Noguchi, {\tt arXiv:hep-lat/0902022}.
  
\bibitem{Gattringer:2000ja} C.~Gattringer and I.~Hip,
  Phys.\ Lett.\ B {\bf 480}, 112 (2000); \\
  C.~Gattringer,
  Phys.\ Rev.\ D {\bf 63}, 114501 (2001); \\
  C.~Gattringer, I.~Hip and C.~B.~Lang, Nucl.\ Phys.\ B {\bf 597}, 451
  (2001).
  
\bibitem{Hasenfratz:1998ri} P.~Hasenfratz, V.~Laliena and
  F.~Niedermayer, Phys.\ Lett.\ B {\bf 427}, 125 (1998).
  
\bibitem{Hasenfratz:1998bb} P.~Hasenfratz, Nucl.\ Phys.\ B {\bf 525},
  401 (1998).
  
\bibitem{Luscher:1998pq} M.~L\"uscher, Phys.\ Lett.\ B {\bf 428}, 342
  (1998).
  
\bibitem{Niedermayer:1998bi} F.~Niedermayer,
  Nucl.\ Phys.\ Proc.\ Suppl.\  {\bf 73}, 105 (1999);\\
  H.~Neuberger,
  Nucl.\ Phys.\ Proc.\ Suppl.\  {\bf 83-84}, 67 (2000);\\
  M.~L\"uscher, Nucl.\ Phys.\ Proc.\ Suppl.\ {\bf 83-84}, 34 (2000).
  
\bibitem{Jansen:2001fn} K.~Jansen,
  Nucl.\ Phys.\ Proc.\ Suppl.\  {\bf 106}, 191 (2002); \\
  P.~Hernandez, Nucl.\ Phys.\ Proc.\ Suppl.\ {\bf 106}, 80 (2002).
  
\bibitem{Knechtli:2000ku} F.~Knechtli and A.~Hasenfratz, Phys.\ Rev.\ 
  D {\bf 63}, 114502 (2001);
  Nucl.\ Phys.\ Proc.\ Suppl.\ {\bf 106}, 1058 (2002).
  
\bibitem{Hasenfratz:gu} P.~Hasenfratz, Prog.\ Theor.\ Phys.\ Suppl.\ 
  {\bf 131}, 189 (1998).
  
\bibitem{Rufenacht:2001pi} F.~Niedermayer, P.~R\"ufenacht and
  U.~Wenger, Nucl.\ Phys.\ B {\bf 597}, 413 (2001);
  Nucl.\ Phys.\ Proc.\ Suppl.\  {\bf 94}, 636 (2001); \\
  P.~R\"ufenacht and U.~Wenger, Nucl.\ Phys.\ B {\bf 616}, 163 (2001).
  
\bibitem{Blatter:1994hy} T.~DeGrand, A.~Hasenfratz and T.~G.~Kovacs,
  Nucl.\ Phys.\ B {\bf 505}, 417 (1997); \\
  T.~DeGrand, A.~Hasenfratz and D.~C.~Zhu, Nucl.\ Phys.\ B {\bf 475},
  321 (1996);
  Nucl.\ Phys.\ B {\bf 478}, 349 (1996); \\
  C.~B.~Lang and T.~K.~Pany,
  Nucl.\ Phys.\ B {\bf 513}, 645 (1998); \\
  T.~Bhattacharya, R.~Gupta and W.~J.~Lee,
  Nucl.\ Phys.\ Proc.\ Suppl.\  {\bf 83}, 860 (2000); \\
  F.~Farchioni, I.~Hip, C.~B.~Lang and M.~Wohlgenannt,
  Nucl.\ Phys.\ Proc.\ Suppl.\  {\bf 73}, 939 (1999); \\
  W.~Bietenholz and U.~J.~Wiese, Nucl.\ Phys.\ B {\bf 464}, 319
  (1996);
  Phys.\ Lett.\ B {\bf 378}, 222 (1996); \\
  F.~Farchioni and V.~Laliena, Nucl.\ Phys.\ B {\bf 521}, 337 (1998);
  Phys.\ Rev.\ D {\bf 58}, 054501 (1998); \\
  W.~Bietenholz and H.~Dilger, Nucl.\ Phys.\ B {\bf 549}, 335 (1999).
  
\bibitem{Hauswirth2002} S.~Hauswirth, PhD thesis, 
{\tt arXiv:hep-lat/0204015}.

\bibitem{Hasenfratz:2000xz} P.~Hasenfratz, S.~Hauswirth, K.~Holland,
  T.~Jorg, F.~Niedermayer and U.~Wenger, Int.\ J.\ Mod.\ Phys.\ C {\bf
    12}, 691 (2001);
  Nucl.\ Phys.\ Proc.\ Suppl.\ {\bf 94}, 627 (2001).
  
\bibitem{Hasenfratz:2001hr} P.~Hasenfratz, S.~Hauswirth, K.~Holland,
  T.~Jorg and F.~Niedermayer, Nucl.\ Phys.\ Proc.\ Suppl.\ {\bf 106},
  799 (2002).
  
\bibitem{Hasenfratz:2001qp} P.~Hasenfratz, S.~Hauswirth, K.~Holland,
  T.~Jorg and F.~Niedermayer, Nucl.\ Phys.\ Proc.\ Suppl.\ {\bf 106},
  751 (2002).
  
\bibitem{Horvath:1998dd} I.~Horvath, Phys.\ Rev.\ Lett.\ {\bf 81},
  4063 (1998);
  W.~Bietenholz, {\tt arXiv:hep-lat/9901005}.
  
\bibitem{Bietenholz:2000iy} W.~Bietenholz, 
 {\tt arXiv:hep-lat/0007017}.
  
\bibitem{Hernandez:1998et} P.~Hernandez, K.~Jansen and M.~L\"uscher,
  Nucl.\ Phys.\ B {\bf 552}, 363 (1999).
  
\bibitem{Bietenholz:2001nu} W.~Bietenholz, I.~Hip and K.~Schilling,
  Nucl.\ Phys.\ Proc.\ Suppl.\ {\bf 106}, 829 (2002); \\
 W.~Bietenholz, {\tt arXiv:hep-lat/0204016}.

\bibitem{Gottlieb:2001sy} S.~Gottlieb, Comput.\ Phys.\ Commun.\ {\bf
    142}, 43 (2001).
  
\bibitem{Sommer:1993ce} R.~Sommer, Nucl.\ Phys.\ B {\bf 411}, 839
  (1994).
  
\bibitem{DeForcrand:1989aa} P.~de Forcrand and R.~Gupta, Nucl.\ Phys.\ 
  Proc.\ Suppl.\ {\bf 9}, 516, (1989).
  
\bibitem{Jegerlehner:1997rn} B.~Jegerlehner, Nucl.\ Phys.\ Proc.\ 
  Suppl.\ {\bf 63}, 958 (1998),
  {\tt arXiv:hep-lat/9612014}.
  
\bibitem{AliKhan:2001tx} A.~Ali Khan {\it et al.}  [CP-PACS
  Collaboration], Phys.\ Rev.\ D {\bf 65}, 054505 (2002).

  
\bibitem{Blum:2000kn} T.~Blum {\it et al.},
  {\tt arXiv:hep-lat/0007038};\\
  S.~J.~Dong {\it et al.},
  {\tt arXiv:hep-lat/0108020};\\
  S.~J.~Dong {\it et al.}, {\tt arXiv:hep-lat/0110044}.
  
\bibitem{DeGrand:2000gq} T.~DeGrand and A.~Hasenfratz, Phys.\ Rev.\ D
  {\bf 64}, 034512 (2001).
  
\bibitem{Bochicchio1985} M.~Bochicchio, L.~Maiani, G.~Martinelli, 
G.~C.~Rossi and M.~Testa, 
Nucl.\ Phys.\ {\bf 262}, 331 (1985).

\bibitem{Giusti1999}  L.~Giusti, F.~Rapuano, M.~Talevi and
  A.~Vladikas, Nucl.\ Phys.\ {\bf 538}, 249 (1999),
  {\tt arXiv:hep-lat/9807014}.

\bibitem{Giusti:2001pk} L.~Giusti, C.~Hoelbling and C.~Rebbi, Phys.\ 
  Rev.\ D {\bf 64}, 114508 (2001).
  
\bibitem{Lee:1997bq} F.~X.~Lee and D.~B.~Leinweber, Phys.\ Rev.\ D
  {\bf 59}, 074504 (1999).
  
\bibitem{Gasser:1983yg} J.~Gasser and H.~Leutwyler, Annals Phys.\ {\bf
    158}, 142 (1984);
  Nucl.\ Phys.\ B {\bf 250}, 465 (1985).
  
\bibitem{Sharpe:1992ft} S.~R.~Sharpe,
  Phys.\ Rev.\ D {\bf 46}, 3146 (1992); \\
  C.~W.~Bernard and M.~F.~Golterman, Phys.\ Rev.\ D {\bf 46}, 853
  (1992).
  
\bibitem{Osborn:1998qb} J.~C.~Osborn, D.~Toublan and
  J.~J.~Verbaarschot,
  Nucl.\ Phys.\ B {\bf 540}, 317 (1999); \\
  P.~H.~Damgaard, J.~C.~Osborn, D.~Toublan and J.~J.~Verbaarschot,
  Nucl.\ Phys.\ B {\bf 547}, 305 (1999); \\
  D.~Toublan and J.~J.~Verbaarschot,
  Nucl.\ Phys.\ B {\bf 560}, 259 (1999); \\
  P.~H.~Damgaard, Nucl.\ Phys.\ Proc.\ Suppl.\ {\bf 106}, 29 (2002).
  
\bibitem{Kiskis:2001zt} J.~E.~Kiskis and R.~Narayanan, Phys.\ Rev.\ D
  {\bf 64}, 117502 (2001).
  
\bibitem{Dong:1993pk} S.~J.~Dong and K.~F.~Liu, Phys.\ Lett.\ B {\bf
    328}, 130 (1994).
  
\bibitem{Edwards:1998hh} R.~G.~Edwards, U.~M.~Heller and R.~Narayanan,
  Phys.\ Rev.\ D {\bf 59}, 091510 (1999).
  
\bibitem{Hernandez:1999gg} P.~Hernandez, K.~Jansen and L.~Lellouch,
  Phys.\ Lett.\ B {\bf 469}, 198 (1999).
  
\bibitem{Sorenson:1992} D.~C.~Sorenson,
  SIAM J.\ Matrix Anal.\ Appl.\ {\bf 13}, 357 (1992); \\
  R.~B.~Lehoucq, D.~C.~Sorenson and C.~Yang, ARPACK Users' Guide,
  SIAM, New York, 1998.
  
\bibitem{AliKhan:2001ym} M.~Teper,
  Nucl.\ Phys.\ Proc.\ Suppl.\  {\bf 83-84}, 146 (2000);\\
  A.~Ali Khan {\it et al.}  [CP-PACS Collaboration],
  Phys.\ Rev.\ D {\bf 64}, 114501 (2001); \\
  T.~G.~Kovacs,
  Nucl.\ Phys.\ Proc.\ Suppl.\  {\bf 106}, 578 (2002); \\
  A.~Hasenfratz,
  Phys.\ Rev.\ D {\bf 64}, 074503 (2001); \\
  G.~S.~Bali {\it et al.}  [SESAM Collaboration], Phys.\ Rev.\ D {\bf
    64}, 054502 (2001).
  
\bibitem{Gattringer:2002mr} C.~Gattringer, R.~Hoffmann and
  S.~Schaefer, arXiv:hep-lat/0203013.
  
\bibitem{Garden:1999fg} J.~Garden, J.~Heitger, R.~Sommer and H.~Wittig
  [ALPHA Collaboration], Nucl.\ Phys.\ B {\bf 571}, 237 (2000).
  
\bibitem{Hernandez:2001yn} P.~Hernandez, K.~Jansen, L.~Lellouch and
  H.~Wittig, JHEP {\bf 0107}, 018 (2001);
  Nucl.\ Phys.\ Proc.\ Suppl.\ {\bf 106}, 766 (2002).
  
\bibitem{Capitani:1998mq} S.~Capitani, M.~L\"uscher, R.~Sommer and
  H.~Wittig [ALPHA Collaboration], Nucl.\ Phys.\ B {\bf 544}, 669
  (1999).
  
\bibitem{DeGrand:2001ie} T.~DeGrand [MILC Collaboration], Phys.\ Rev.\ 
  D {\bf 64}, 117501 (2001);
  Phys.\ Rev.\ D {\bf 63}, 034503 (2001).
  
\bibitem{Nishigaki:1998is} S.~M.~Nishigaki, P.~H.~Damgaard and
  T.~Wettig,
  Phys.\ Rev.\ D {\bf 58}, 087704 (1998); \\
  P.~H.~Damgaard and S.~M.~Nishigaki, Phys.\ Rev.\ D {\bf 63}, 045012
  (2001).
  
\bibitem{Banks:1979yr} T.~Banks and A.~Casher, Nucl.\ Phys.\ B {\bf
    169}, 103 (1980).
  
\bibitem{Horvath:2001ir} I.~Horvath, N.~Isgur, J.~McCune and
  H.~B.~Thacker, Phys.\ Rev.\ D {\bf 65}, 014502 (2002).
  
\bibitem{DeGrand:2001pj} T.~DeGrand and A.~Hasenfratz,
  Phys.\ Rev.\ D {\bf 65}, 014503 (2002); \\
  C.~Gattringer, M.~G\"ockeler, P.~E.~Rakow, S.~Schaefer and
  A.~Sch\"afer, Nucl.\ Phys.\ B {\bf 618}, 205 (2001);
  Nucl.\ Phys.\ B {\bf 617}, 101 (2001); \\
  T.~Blum {\it et al.},
  Phys.\ Rev.\ D {\bf 65}, 014504 (2002); \\
  R.~G.~Edwards and U.~M.~Heller,
  Phys.\ Rev.\ D {\bf 65}, 014505 (2002); \\
  I.~Hip, T.~Lippert, H.~Neff, K.~Schilling and W.~Schroers, Phys.\ 
  Rev.\ D {\bf 65}, 014506 (2002).

\bibitem{Horvath} S.~J.~Dong {\it et al.},
Nucl.\ Phys.\ Proc.\ Suppl.\ {\bf 106}, 563 (2002); 
 I.~Horvath {\it et al.}, {\tt
    arXiv:hep-lat/0201008};
 N.~Cundy, M.~Teper and U.~Wenger, {\tt
    arXiv:hep-lat/0203030}.
  
\bibitem{Gattringer:2002gn} C.~Gattringer, {\tt
    arXiv:hep-lat/0202002}.
  
\bibitem{Kikukawa:1998bg} Y.~Kikukawa and A.~Yamada, {\tt
    arXiv:hep-lat/9810024}.
  
\bibitem{Narayanan:1998uu} R.~Narayanan, Phys.\ Rev.\ D {\bf 58},
  97501 (1998).
  
\bibitem{Chandrasekharan:1999wg} S.~Chandrasekharan,, Phys.\ Rev.\ D
  {\bf 60}, 074503 (1999).
  
\bibitem{Hernandez:2000iw} P.~Hernandez, K.~Jansen and M.~L\"uscher,
  {\tt arXiv:hep-lat/0007015}.

\bibitem{DeGrand2002} T.~DeGrand and U.~Heller,
  {\tt arXiv:hep-lat/0202001}.

\bibitem{Edwards:1998vu}
R.~G.~Edwards, U.~M.~Heller and R.~Narayanan,
Nucl.\ Phys.\ Proc.\ Suppl.\  {\bf 73} (1999) 500
[arXiv:hep-lat/9810019].

\bibitem{DeGrand2000} T.~DeGrand, Phys.\ Rev.\ D {\bf 63}, 034503 (2001).
  {\tt arXiv:hep-lat/0007046}.

\end{thebibliography}
\end{document}